\documentclass[12pt]{article}

\font\grande=cmr9.5 scaled \magstep4
\font\medio=cmr9.5 scaled \magstep2
\outer\def\beginsection#1\par{\medbreak\bigskip
      \message{#1}\leftline{\bf#1}\nobreak\medskip
\vskip-\parskip
      \noindent}
\usepackage{graphicx} 
\begin{document}
\bibliographystyle {unsrt}

\titlepage

\begin{flushright}
CERN-PH-TH/2014-119
\end{flushright}

\vspace{10mm}
\begin{center}
{\grande Magnetization of fluid phonons}\\
\vspace{4mm}
{\grande and large-scale curvature perturbations}\\
\vspace{1.5cm}
 Massimo Giovannini
 \footnote{Electronic address: massimo.giovannini@cern.ch}\\
\vspace{1cm}
{{\sl Department of Physics, 
Theory Division, CERN, 1211 Geneva 23, Switzerland }}\\
\vspace{0.5cm}
{{\sl INFN, Section of Milan-Bicocca, 20126 Milan, Italy}}
\vspace*{0.5cm}
\end{center}

\vskip 0.5cm
\centerline{\medio  Abstract}
The quasinormal mode of a gravitating and magnetized fluid in a spatially 
flat, isotropic and homogeneous cosmological background is derived in the presence of 
the fluid sources of anisotropic stress and of the entropic fluctuations of the plasma.
The obtained gauge-invariant description involves a system of two coupled differential equations whose 
physical content is analyzed in all the most relevant situations. 
The Cauchy problem of large-scale curvature perturbations during the radiation dominated stage of expansion
can be neatly formulated and its general solution is shown to depend on five initial data 
assigned when the relevant physical wavelengths are larger than the particle horizon. The 
consequences of this approach are explored.
\vskip 0.5cm

\noindent

\vspace{5mm}

\vfill
\newpage
\renewcommand{\theequation}{1.\arabic{equation}}
\setcounter{equation}{0}
\section{Formulation of the problem and basic results}
\label{sec1}
The presumed existence of large-scale magnetic fields prior to matter-radiation equality (see e.g. \cite{zeldovich,mg1})
 affects the evolution of all the scalar inhomogeneities of the plasma. 
 The discussion of this physical system is feasible but algebraically cumbersome. 
Part of the technical complication stems from the scalar 
nature of the problem involving, on equal footing, the curvature perturbations, the diverse sources of 
anisotropic stress and the entropic fluctuations of the medium.
In this investigation we ought to propose a simpler but effective framework for solving the evolution and normalization of the magnetized curvature 
perturbations. This strategy shall also be applied to the generalized analysis of the Cauchy problem for large-scale curvature 
perturbations.

In a conformally flat and homogeneous background geometry the fluctuations of a gravitating, irrotational and relativistic fluid admit a normal mode
that has been originally obtained and scrutinized by Lukash \cite{lukash1} (see also \cite{lukash2}) even prior to the actual formulation of the conventional inflationary paradigm and in the context of the pioneering analyses of the relativistic theory of large-scale inhomogeneities \cite{lif,lif1}. 
If the gravitating fluid contains arbitrary sources of anisotropic stress and of entropy 
perturbations the system admits a {\em quasinormal mode} 
that reduces to the Lukash variable in the absence of all the supplementary contributions. The magnetized quasinormal 
mode turns out to be coupled to the evolution of the anisotropic stress. The two resulting equations form a self-contained system 
that shall be derived and solved hereunder in various physical situations.  

In anticipation of the actual derivation, the evolution equation of the quasinormal mode can be written\footnote{We shall use the standard notations: the prime denotes the derivation with respect to the conformal time coordinate $\tau$; $\rho_{t}$ and $p_{t}$ denote the total energy density and pressure of the system, $a$ is the scale factor of the conformally flat metric in four space-time dimensions and ${\mathcal H} = a'/a = a H$ where 
$H$ is the Hubble rate. Note that  $c_{\mathrm{st}}$ is the {\em total} sound speed of the system. Further notational precisions can be found in sections \ref{sec2} and \ref{sec3}.}, in a globally neutral plasma, as:
\begin{equation}
{\mathcal R}'' + 2 \frac{z_{t}'}{z_{t}} {\mathcal R}' - c_{\mathrm{st}}^2 \nabla^2 {\mathcal R} = {\mathcal S}_{{\mathcal R}}[z_{t},\,c_{\mathrm{st}}^2; \delta_{s}\rho_{B},\, \delta_{s}\rho_{E},\, \Pi_{t},\,\delta p_{\mathrm{nad}}],
\label{Rfirst}
\end{equation}
where ${\mathcal S}_{{\mathcal R}}[.\,.\,.]$ denotes the functional of the two
{\em homogeneous} background fields $z_{t}(\tau)$ and $c_{\mathrm{t}}^2(\tau)$
 \begin{equation}
z_{t} = \frac{a^2 \sqrt{p_{t} + \rho_{t}}}{{\mathcal H} c_{\mathrm{st}}}, \qquad c_{\mathrm{st}}^2 = \frac{p_{t}'}{\rho_{t}'},
\label{Rsecond}
\end{equation}
 and of three {\em inhomogeneous background}  fields, namely: {\it (i)}  the total anisotropic stress $\Pi_{t}$;
{\it (ii)}  the electromagnetic energy density (denoted respectively by $\delta \rho_{B}$ and $\delta \rho_{E}$);
{\it (iii)} the non-adiabatic fluctuation of the total pressure (i.e. $\delta p_{\mathrm{and}}$). 

The limit ${\mathcal S}_{{\mathcal R}}[.\,.\,.] \to 0$ in Eq. (\ref{Rfirst}) reproduces exactly the equation obeyed by the Lukash variable ${\mathcal R}$ (or $z_{t} \,{\mathcal R}$ as defined in \cite{lukash1,lukash2}). This variable coincides, in fact, with the curvature perturbation on comoving orthogonal hypersurfaces and it is invariant under infinitesimal coordinate transformations as required in the context of the Bardeen formalism \cite{bard1}. In the absence of magnetic fields, subsequent analyses  \cite{KS,chibisov} followed the same logic of \cite{lukash1} but in the case of scalar field matter.  All the normal modes identified in \cite{lukash1,lukash2,KS,chibisov}  are related to the (rescaled) curvature perturbations on comoving orthogonal hypersurfaces \cite{br1,bard2}. 

({\it i}) The first of the three inhomogeneous contributions appearing in Eq. (\ref{Rfirst}) and mentioned after Eq. (\ref{Rsecond}) is 
the total anisotropic stress which is given as the sum of all the anisotropic stresses of the system, namely the fluid and the electromagnetic parts:
\begin{equation}
\Pi_{t}(\vec{x}, \tau)= \Pi_{f}(\vec{x}, \tau) + \Pi_{B}(\vec{x}, \tau) + \Pi_{E}(\vec{x}, \tau).
\label{Rfifth}
\end{equation}

{\it (ii)} The second inhomogeneous contribution is given by the electromagnetic inhomogeneities\footnote{Note that $\Pi_{ij}^{(B)}$ and  $\Pi_{ij}^{(E)}$ 
 are defined in the standard way, namely  $\Pi_{ij}^{(B)}=  [B_{i}\,B_{j}- B^2(\vec{x},\tau)\delta_{ij}/3]/(4 \pi a^4)$ and 
 $\Pi_{ij}^{(E)}= [E_{i}\,E_{j}- E^2(\vec{x},\tau)\delta_{ij}/3]/(4 \pi a^4)$.}
\begin{eqnarray}
&& \delta_{s} \rho_{B}(\vec{x},\tau) = \frac{B^2(\vec{x},\tau)}{4 \pi a^4}, \qquad \delta_{s} \rho_{E}(\vec{x},\tau) = \frac{E^2(\vec{x},\tau)}{4 \pi a^4},
\label{Rsixtha}\\
&&\nabla^2 \Pi_{B}(\vec{x},\tau) = \partial_{i} \partial_{j} \Pi^{ij}_{(B)}(\vec{x},\tau), \qquad 
\nabla^2 \Pi_{E}(\vec{x},\tau) = \partial_{i} \partial_{j} \Pi^{ij}_{(E)}(\vec{x},\tau).
\label{Rsixthd}
\end{eqnarray}
For practical reasons it is convenient to introduce $\sigma_{E}$ and $\sigma_{B}$, namely the dimensionless counterpart of $\Pi_{E}$ and $\Pi_{B}$: 
\begin{equation}
\Pi_{B}(\vec{x},\tau) = (p_{\gamma} + \rho_{\gamma}) \sigma_{B}(\vec{x},\tau), \qquad \Pi_{E}(\vec{x},\tau) = (p_{\gamma} + \rho_{\gamma}) \sigma_{E}(\vec{x},\tau),
\label{Rsixthe}
\end{equation}
where $p_{\gamma}$ and $\rho_{\gamma}$ are the pressures and energy density of the photon background.

{\it (iii)} Finally the third inhomogeneous contribution the source term appearing in Eq. (\ref{Rfirst}) is the non-adiabatic pressure 
fluctuation $\delta p_{\mathrm{nad}}$ (see e.g. \cite{hh0,hh1}) that can be written\footnote{Within the 
notation of Eq. (\ref{DPNAD1}) the inhomogeneities of the total pressure are given
by $\delta_{s} p_{t} = c_{\mathrm{st}}^2 \, \delta_{s} \rho_{t} + \delta p_{\mathrm{nad}}$ where $\delta_{s}$ shall denote throughout the paper the scalar component 
of the corresponding quantity.} as:
\begin{equation}
\delta p_{\mathrm{nad}}(\vec{x},\tau) = \sum_{\mathrm{ij}}\frac{\partial p_{\mathrm{t}}}{\partial \varsigma_{\mathrm{ij}}} \delta \varsigma_{\mathrm{ij}}(\vec{x},\tau)
 = \frac{1}{6 {\mathcal H} \rho_{\mathrm{t}}'} \sum_{\mathrm{ij}} \rho_{\mathrm{i}}' \rho_{\mathrm{j}}' 
(c_{\mathrm{si}}^2 - c_{\mathrm{sj}}^2) {\mathcal S}_{\mathrm{ij}}(\vec{x},\tau),\qquad 
{\mathcal S}_{\mathrm{ij}}(\vec{x},\tau) = \frac{\delta \varsigma_{\mathrm{ij}}(\vec{x},\tau)}{\varsigma_{\mathrm{ij}}},
\label{DPNAD1}
\end{equation}
where the indices i and j are not tensor indices but denote 
two generic species of the pre-equality plasma.
Furthermore, in Eq. (\ref{DPNAD1}), $c_{\mathrm{si}}^2$ and $c_{\mathrm{sj}}^2$ are the sound speeds of two (generic) species 
of the plasma; $\delta \varsigma_{\mathrm{ij}}(\vec{x},\tau)$ is the fluctuation of the specific entropy computed for a given pair of species and 
${\mathcal S}_{\mathrm{ij}}(\vec{x},\tau)$, as indicated, is the relative fluctuation 
of $\varsigma_{\mathrm{ij}}$. 
With the precisions of Eqs. (\ref{Rfifth}), (\ref{Rsixtha})--(\ref{Rsixthe}) and (\ref{DPNAD1}) the general form of ${\mathcal S}_{{\mathcal R}}(\vec{x},\tau)$ can be expressed as
\begin{equation}
{\mathcal S}_{{\mathcal R}}(\vec{x},\tau) =  \Sigma_{\mathcal R}' + 2 \frac{z_{t}'}{z_{t}} \Sigma_{\mathcal R}+ \frac{ 3 a^{4}}{z_{t}^2 } \Pi_{t},
\label{Rthird}
\end{equation}
where the auxiliary quantity $\Sigma_{\mathcal R}(\vec{x},\tau)$ is given by:
\begin{equation}
\Sigma_{{\mathcal R}}(\vec{x},\tau) = - \frac{{\mathcal H}}{p_{t} + \rho_{t}} \delta p_{\mathrm{nad}} + \frac{{\mathcal H}}{p_{t} + \rho_{t}} \biggl[ \biggl( c_{\mathrm{st}}^2 - \frac{1}{3}\biggr)(\delta_{s}\rho_{E} + \delta_{s}\rho_{B})  + \Pi_{t} \biggr].
\label{Rsixth}
\end{equation}

The result of Eq. (\ref{Rfirst}) must be complemented by an equation for the anisotropic stress that only couples 
to the quasinormal mode. Consider, in this respect, the standard $\Lambda$CDM paradigm (where $\Lambda$ qualifies the dark energy 
component and CDM stands for the cold dark matter contribution). In this case the anisotropic stress 
of the fluid comes only from the neutrino sector\footnote{The strategy examined in this paper applies also to more general situations. However, for the 
sake of concreteness we shall focus on the situation where the main source of the anisotropic stress of the fluid comes from massless neutrinos, as 
demanded by the $\Lambda$CDM paradigm. Of course due to the presence of the electromagnetic degrees of freedom $\Pi_{f} \neq \Pi_{t}$.}. Defining therefore $\Pi_{f} = (p_{\nu} + \rho_{\nu}) \sigma_{\nu}$ in Eq. (\ref{Rsixth}) the evolution equation of $\sigma_{\nu}$ becomes
\begin{equation}
\sigma_{\nu}''' + \frac{8}{5} {\mathcal H}^2 R_{\nu} \Omega_{R} \sigma_{\nu}' - \frac{6}{7} \nabla^2 \sigma_{\nu}' - \frac{32 a^4 {\mathcal H}}{5 c_{\mathrm{st}}^2 z_{t}^2} \Pi_{t}
= \frac{4 z_{t}^2 }{15 \overline{M}_{\mathrm{P}}^2}  \biggl[ \biggl(\frac{{\mathcal H}}{a^2}\biggr)'  ({\mathcal R}' - \Sigma_{{\mathcal R}}) +  c_{\mathrm{st}}^2 \biggl(\frac{{\mathcal H}}{a^2}\biggr)\nabla^2 {\mathcal R} \biggr]
\label{Rseventh}
\end{equation}
where $\Omega_{R} = \rho_{R}/\rho_{t}$ is the critical fraction of radiation; $R_{\nu}$ and $R_{\gamma}$ count the fraction of neutrinos and 
photons in the radiation plasma. Once $\delta_{s}\rho_{B}$ and $\Pi_{B}$ are given, Eqs. (\ref{Rfirst}) and (\ref{Rseventh}) form a closed and gauge-invariant system of equations that can be solved under various approximations.

The system of Eqs. (\ref{Rfirst}) and (\ref{Rseventh}) bears some analogy with the inhomogeneous evolution of two minimally coupled scalar 
fields in a conformally flat and homogeneous background: as in our case the dynamics is described 
in terms of two quasinormal mode that are promoted to the status of exact canonical variables when one of the two fields is absent (see, e.g. \cite{h0,h1,h2} and discussions therein). Here the situation is similar but also rather different insofar as Eqs. (\ref{Rfirst}) and  (\ref{Rseventh}) are, respectively, 
second and third order partial differential equations. The present approach simplifies and improves former analytical and numerical discussions.
For instance it was pointed out in \cite{mm1} that large-scale magnetic fields affect the scalar modes of the plasma, modify the initial conditions of the Einstein-Boltzmann hierarchy  and ultimately change the temperature and polarization anisotropies. The results of \cite{mm1} motivate the actual determination of the magnetized
temperature and polarization anisotropies \cite{mm2}.  After Refs. \cite{mm1,mm2} different groups pursued similar analyses (see \cite{mm3} and \cite{mm4} for an incomplete list of references). The present construction may also some relevance for neighbouring problems where the fluctuations of the anisotropic stress play a physical role \cite{anis1}. 

This paper is organized as follows.
In section \ref{sec2}  the synchronous gauge derivation will be presented.  In section \ref{sec3} 
the main results are derived along the perspective of the longitudinal gauge.  The overall symmetries of the system are examined in section \ref{sec4}.
Various analytic solutions are discussed in section \ref{sec5}.  In section \ref{sec6}  the formalism is applied to the analysis of the Cauchy problem for large-scale curvature perturbations. The concluding remarks are collected in section \ref{sec7}.
 
\renewcommand{\theequation}{2.\arabic{equation}}
\setcounter{equation}{0}
\section{Synchronous gauge derivation}
\label{sec2}
We shall consider a spatially flat Friedmann-Robertson-Walker background 
in the conformal time parametrization; as already mentioned in section \ref{sec1}, the total energy density and pressure 
of the plasma shall be denoted by  $\rho_{t}$ and $p_{t}$. With these conventions the evolution equations  of the homogeneous 
background are:
\begin{eqnarray}
&& {\mathcal H}^2 = \frac{8 \pi G}{3} a^2 \rho_{t},
\label{FL1}\\
&& {\mathcal H}^2 - {\mathcal H}' = 4 \pi G a^2 (p_{t} + \rho_{t}), 
\label{FL2}\\
&& \rho_{t}' + 3 {\mathcal H} (\rho_{t} + p_{t}) =0,
\label{FL3}
\end{eqnarray}
where ${\mathcal H} = a'/a$; the Planck mass is defined as  $8 \pi G = 1/\overline{M}_{\mathrm{P}}^2$. In the 
synchronous gauge\footnote{We remind that, in this gauge, the non-vanishing 
entries of the perturbed metric are given, in Fourier space, by $\delta_{s}g_{ij}(k,\tau) = a^2(\tau)[ \hat{k}_{i} \hat{k}_{j} h(k,\tau) + 6 \xi(k,\tau) (\hat{k}_{i} \hat{k}_{j} - \delta_{ij}/3)]$ where $\hat{k}_{i} = k_{i}/|\vec{k}|$; as mentioned after Eq. (\ref{DPNAD1}) $\delta_{s}$ denotes the scalar mode of the fluctuation of the metric. } the metric fluctuations are related to the anisotropic 
stress and to the curvature perturbations by the following relevant pair of equations \cite{mm1,mm2}:
\begin{eqnarray}
&& {\mathcal R}' = \Sigma_{R} - \frac{a^2 k^2 \xi}{4 \pi G {\mathcal H} z_{t}^2} + \frac{a^2 (h + 6 \xi)'}{8 \pi G z_{t}^2},
\label{SYN1}\\
&& (h + 6 \xi)'' + 2 {\mathcal H} (h + 6 \xi)' - 2 k^2 \xi = 24 \pi G a^2 \Pi_{t},
\label{SYN2}
\end{eqnarray}
where $\Pi_{t}$ is the total anisotropic stress (already introduced in Eq. (\ref{Rfifth})) that will now be explicitly rewritten as
\begin{equation}
 \Pi_{t} = \biggl[ (p_{\nu} + \rho_{\nu}) \sigma_{\nu} + (p_{\gamma}+ \rho_{\gamma}) \sigma_{B} \biggr].
\label{SYN2a}
\end{equation}
In Eq. (\ref{SYN2a}) the only source of fluid anisotropic stress has been identified with the neutrinos that will be taken to be 
massless as in the vanilla $\Lambda$CDM paradigm (see e.g. \cite{weinbook,mgbook} and discussions therein).

\subsection{Evolution of the quasinormal mode}

Moving $\Sigma_{{\mathcal R}}$ at the left hand side of Eq. (\ref{SYN1}) and taking the first time derivative of both sides of the obtained 
equation, the following result can easily be derived:
\begin{equation}
({\mathcal R}' - \Sigma_{{\mathcal R}})' + 2 \frac{z_{t}'}{z_{t}} ({\mathcal R}' - \Sigma_{{\mathcal R}})= \frac{3 a^4 \Pi_{t}}{z_{t}^2} - \frac{k^2 a^2 \xi}{4\pi G z_{t}^2} 
- \frac{a^2 k^2 \xi'}{4 \pi G {\mathcal H} z_{t}^2} + \frac{a^2 k^2 {\mathcal H}' \xi}{4 \pi G {\mathcal H}^2 z_{t}^2}.
\label{SYN3}
\end{equation}
In Eq. (\ref{SYN3}) the expressions containing $(h + 6 \xi)''$ and $(h+ 6\xi)'$ (arising as a result of the explicit derivation of both sides 
of Eq. (\ref{SYN1})) have been eliminated by using Eq. (\ref{SYN2}) in combination with Eq. (\ref{SYN1}).  
Recalling then the synchronous gauge expression 
for the curvature perturbations on comoving orthogonal hypersurfaces, i.e. 
\begin{equation}
{\mathcal R} = \xi + \frac{{\mathcal H}\, \xi'}{{\mathcal H}^2 - {\mathcal H}'}, 
\label{SYN4}
\end{equation}
Eq. (\ref{SYN3}) can be expressed, after some algebraic manipulations, as:
\begin{equation}
({\mathcal R}' - \Sigma_{{\mathcal R}})' + 2 \frac{z_{t}'}{z_{t}} ({\mathcal R}' - \Sigma_{{\mathcal R}}) + c_{\mathrm{st}}^2 k^2 {\mathcal R}= \frac{3 a^4}{z_{t}^2} \Pi_{t},
\label{SYN5}
\end{equation}
which is nothing but the result anticipated in Eqs. (\ref{Rfirst}) (see also Eqs. (\ref{Rthird})--(\ref{Rsixth})) and now explicitly derived in Fourier space.  

\subsection{Evolution of the anisotropic stress}
The equation for $\sigma_{\nu}$ can be written, in the synchronous gauge, as
\begin{equation}
\sigma_{\nu}' = \frac{4}{15} \theta_{\nu} - \frac{3}{10} k {\mathcal F}_{\nu\, 3} - \frac{2}{15} (h + 6 \xi)',
\label{SYN6}
\end{equation}
where, following the standard conventions (see e.g. \cite{mm1,mm2}), $\theta_{\nu}$ and $\delta_{\nu}$ are, respectively, the divergence 
of the three-velocity of the neutrinos and the corresponding density contrast all computed in the synchronous gauge; ${\mathcal F}_{\nu\,3}$ is the octupole 
of the neutrino phase-space distribution. After taking the first time derivative of both sides of Eq. (\ref{SYN6}),
the equations of the lower multipoles:
\begin{eqnarray}
&& \delta_{\nu}' = - \frac{4}{3} \theta_{\nu} + \frac{2}{3} h', 
\label{SYN6a}\\
&& \theta_{\nu}' = \frac{k^2}{4} \delta_{\nu} - k^2 \sigma_{\nu},
\label{SYN6b}\\
&& {\mathcal F}_{\nu\ell}'= \frac{k}{2\ell + 1} [ \ell {\mathcal F}_{\nu(\ell -1)} - (\ell + 1) {\mathcal F}_{\nu(\ell+1)}],\qquad \ell\geq 3,
\label{SYN6c}
\end{eqnarray}
can be used in the obtained result.  We shall now assume\footnote{This assumption is not strictly essential. It corresponds to the standard truncation of the neutrino hierarchy that is commonly posited in the case of the standard adiabatic mode 
\cite{admode} (see also \cite{pee}). Other truncation schemes can be explored but will not be explicitly discussed here and do not 
change the overall spirit of the analysis.} that ${\mathcal F}_{\nu \ell}= 0$ for all $\ell\geq 3$, i.e. ${\mathcal F}_{\nu\, 3}=0$ but, according to Eq. (\ref{SYN6c}), ${\mathcal F}_{\nu 3}' \neq 0$.  Recalling that ${\mathcal F}_{\nu 2} = 2 \sigma_{\nu}$ and using Eqs. (\ref{SYN6b}) and (\ref{SYN6c}), the first derivative of both sides 
of Eq. (\ref{SYN6}) implies:
\begin{equation}
\sigma'' = \frac{k^2}{15} \delta_{\nu} - \frac{11}{21} k^2 \sigma_{\nu} - \frac{2}{15} (h + 6 \xi)''.
\label{SYN7}
\end{equation}
We can take a further time derivative of both sides of Eq. (\ref{SYN7}) and insert Eq. (\ref{SYN6a})  into the obtained expression. Equation 
(\ref{SYN6}) is subsequently used to get rid of $\theta_{\nu}$.
The overall  result of this procedure leads to the following intermediate equation:
\begin{equation}
\sigma_{\nu}^{\prime\prime\prime} + \frac{6}{7} k^2\sigma_{\nu}' + \frac{4}{15} k^2 \, \xi' + \frac{2}{15} ( h + 6 \xi)^{\prime\prime\prime}=0.
\label{SYN8}
\end{equation}
In Eq. (\ref{SYN8}) the term $( h + 6 \xi)^{\prime\prime\prime}$ is eliminated through the relation obtainable from the derivative of both sides of Eq. (\ref{SYN2}). Similarly 
the term $(h + 6\xi)''$  can be replaced by using, again, Eq. (\ref{SYN2}) in its current form. We can finally trade the remaining term (i.e. $(h + 6 \xi)'$ )  for ${\mathcal R}$ and its derivative by inverting Eq. (\ref{SYN1}), i.e. 
\begin{equation}
(h + 6 \xi)' = \frac{8 \pi G  z_{t}^2}{a^2} ({\mathcal R}'  - \Sigma_{{\mathcal R}}) + \frac{2 k^2 \xi}{{\mathcal H}}.
\label{SYN9}
\end{equation}
The final equation we are looking for is thus given by:
\begin{eqnarray}
&& \sigma_{\nu}''' + \frac{8}{5} {\mathcal H}^2 R_{\nu} \Omega_{R} \sigma_{\nu}' + \frac{6}{7} k^2 \sigma_{\nu}' - \frac{32}{5}  {\mathcal H}^3 R_{\nu} \Omega_{R} 
\biggl( \sigma_{\nu} + \frac{R_{\gamma}}{R_{\nu}} \sigma_{B}\biggr)
\nonumber\\
&&= \frac{8}{15 c_{\mathrm{st}}^2} \biggl( {\mathcal H} - \frac{{\mathcal H}'}{{\mathcal H}}\biggr) \biggl(\frac{{\mathcal H}'}{{\mathcal H}} - 2 {\mathcal H}\biggr) ({\mathcal R}' - \Sigma_{{\mathcal R}} ) + \frac{8}{15} \biggl(\frac{{\mathcal H}'}{{\mathcal H}} - {\mathcal H}\biggr) \, k^2 {\mathcal R},
\label{SYN10}
\end{eqnarray}
where $\Omega_{R} = \rho_{R}/\rho_{t}$ is the critical fraction of radiation; as anticipated $R_{\nu}$ and $R_{\gamma}$ count the fraction of neutrinos and photons in the radiation plasma. 

Equation (\ref{SYN10}) can be formally written in different ways using the relations holding among the homogeneous background fields. 
In particular,  from the definition of $z_{t}$ the following simple identity can be easily derived thanks to Eqs. (\ref{FL1}) and (\ref{FL2}):
\begin{equation}
\biggl( {\mathcal H} - \frac{{\mathcal H}'}{{\mathcal H}}\biggr)  = \frac{4 \pi G c_{\mathrm{st}}^2 {\mathcal H}}{a^2} z_{t}^2.
\label{REL1}
\end{equation}
Inserting Eq. (\ref{REL1}) into Eq. (\ref{SYN10}) we obtain the wanted form of the equation which is the Fourier space version of the one 
already mentioned in Eq. (\ref{Rseventh})
\begin{eqnarray}
&& \sigma_{\nu}''' + \frac{8}{5} {\mathcal H}^2 R_{\nu} \Omega_{R} \sigma_{\nu}' + \frac{6}{7} k^2 \sigma_{\nu}' - \frac{32}{5}  {\mathcal H}^3 R_{\nu} \Omega_{R} 
\biggl( \sigma_{\nu} + \frac{R_{\gamma}}{R_{\nu}} \sigma_{B}\biggr) 
\nonumber\\
&& = \frac{32 \pi G}{15} z_{t}^2 \biggl[ \biggl(\frac{{\mathcal H}}{a^2}\biggr)'  ({\mathcal R}' - \Sigma_{{\mathcal R}} )- \frac{k^2 c_{\mathrm{st}}^2{\mathcal H}}{a^2}   {\mathcal R}\biggr],
\label{TWOA}
\end{eqnarray}
recall, in fact, that within the notations established after Eq. (\ref{FL3}) $8 \pi G = 1/\overline{M}_{\mathrm{P}}^2$.

Equations (\ref{SYN5}) and  (\ref{SYN10}) have three relevant properties: they are exact to linear order, they are explicitly invariant under infinitesimal coordinate transformations and they are decoupled from all the remaining equations (even if coupled between them).
To appreciate the importance of the three aforementioned properties it is useful to mention that one 
can easily obtain equations that are neither gauge-invariant nor exact to linear order. 
An example is, for instance, the following equation\footnote{Equation (\ref{SYN11}) follows from Eq. (\ref{SYN6}) if we  assume  that $\sigma_{\nu}' = - 2(h + 6 \xi)'/15$; inserting 
the obtained result into Eq. (\ref{SYN2}) leads to Eq. (\ref{SYN11}) that does not have the same properties of Eq. (\ref{SYN10}): it is neither gauge-invariant
nor exact to linear order.}
\begin{equation}
\sigma_{\nu}'' + 2 {\mathcal H} \sigma_{\nu}' + \frac{4}{15} k^2 \xi + \frac{8}{5} {\mathcal H}^2 \Omega_{R} R_{\nu} \biggl[ \sigma_{\nu} + \frac{R_{\gamma}}{R_{\nu}} \sigma_{B} \biggr] \simeq 0.
\label{SYN11}
\end{equation}
First, in Eq. (\ref{SYN11})  $\sigma_{\nu}$ is gauge-invariant but $\xi$ is not: then the whole 
equation does change for infinitesimal coordinate transformations. Second,
Eq. (\ref{SYN11}) is derived by postulating that  $\theta_{\nu} \ll \sigma_{\nu}'$; the latter requirement is not satisfied by the standard 
adiabatic solution and it is therefore questionable. 
Equation (\ref{SYN11}) and other similar equations are not appropriate for the present approach which rests on the 
validity of Eqs. (\ref{SYN5}) and (\ref{SYN10}).  

We finally remark that the potential ambiguities arising in the synchronous gauge \cite{PV1} (see also \cite{weinberg}) play no role in the present derivation
since we are just using the synchronous coordinate to compute a set of equations that are, ultimately, gauge-invariant. Furthermore, 
as already remarked in the past \cite{mm1,mm2} the synchronous gauge (together with the uniform cirvature gauge \cite{h0,h1}) is 
 more suitable for the description of the modes with potentially relevant anisotropic stresses.

\renewcommand{\theequation}{3.\arabic{equation}}
\setcounter{equation}{0}
\section{Longitudinal gauge derivation}
\label{sec3}
Since Eqs. (\ref{SYN5}) and (\ref{SYN10}) are ultimately gauge-invariant they can be obtained in any other gauge. Let us 
consider, for simplicity, the conformally Newtonian gauge\footnote{We remind that, in this gauge, the non-vanishing 
entries of the perturbed metric are given by $\delta_{s} g_{00}(k,\tau) = 2 a^2(\tau)\,\phi(k,\tau) $ and $\delta_{s}g_{ij}(k,\tau) = 2 a^2(\tau)\, \psi(k,\tau) \delta_{ij}$.}.  While the longitudinal analysis is presented as a useful cross-check, the synchronous description is probably the most suitable, as already stressed at the end of the previous section. 
 
 \subsection{Evolution of the quasinormal mode}

Let us start by recalling that the relation between the curvature perturbations on comoving orthogonal hypersurfaces and the longitudinal degrees of freedom of the metric \cite{bard1} is given by (see e.g. \cite{mm1,mm2} and also \cite{h0,h1,h2} for slightly different notations):
\begin{eqnarray}
&& {\mathcal R} = - \psi - \frac{{\mathcal H}  ( {\mathcal H} \phi + \psi')}{{\mathcal H}^2 - {\mathcal H}'},
\label{LON1}\\
&& {\mathcal R}' = \Sigma_{{\mathcal R}} + \frac{a^2 k^2 \psi}{4\pi G {\mathcal H} z_{t}^2}.
\label{LON2}
\end{eqnarray}
We follow here a procedure that is similar to the one already discussed in section \ref{sec2}. Thus, the equation for the quasinormal mode is obtained by deriving both sides of Eq. (\ref{LON2}) and by using, in the obtained 
expression, Eqs. (\ref{LON1}) and (\ref{LON2}). The result of this manipulation is 
\begin{equation}
{\mathcal R}'' + 2 \frac{z_{t}'}{z_{t}} {\mathcal R}' = \Sigma_{{\mathcal R}}' + 2 \frac{z_{t}'}{z_{t}} \Sigma_{{\mathcal R}} 
- \frac{a^2 k^2 ({\mathcal H}^2 - {\mathcal H}')}{ 4 \pi G {\mathcal H} z_{t}^2} {\mathcal R} + 
\frac{k^2 a^2 {\mathcal H}}{4 \pi G {\mathcal H} z_{t}^2} (\psi - \phi).
\label{LON3}
\end{equation}
In the longitudinal gauge the total anisotropic stress accounts for the mismatch between 
the two longitudinal fluctuations of the metric. In Fourier space we have 
\begin{equation}
k^2 (\phi - \psi) = - 12\pi G a^2 [(p_{\nu} + \rho_{\nu}) \sigma_{\nu} + (p_{\gamma} + \rho_{\gamma}) \sigma_{B} ].
\label{LON5}
\end{equation}
Inserting Eq. (\ref{LON5}) into Eq. (\ref{LON3}) the obtained result coincides, as expected, with Eq. (\ref{SYN5}).

\subsection{Evolution of the anisotropic stress}

Assuming, as in section \ref{sec2}, that the whole anisotropic stress of the fluid 
comes from the neutrino sector, the  lowest multipoles of the neutrino hierarchy read, in the longitudinal gauge,
\begin{eqnarray}
&& \overline{\delta}_{\nu}' = - \frac{4}{3} \overline{\theta}_{\nu} + 4 \psi',
\label{LON5a}\\
&&\overline{\theta}_{\nu}' =  \frac{k^2}{4} \overline{\delta}_{\nu} - k^2 \sigma_{\nu} + k^2 \phi,
\label{LON5b}\\
&& \sigma_{\nu}' = \frac{4}{15} \overline{\theta}_{\nu} - \frac{3}{10} k {\mathcal F}_{\nu 3}.
\label{LON5c}
\end{eqnarray}
In Eqs. (\ref{LON5a}), (\ref{LON5b}) and (\ref{LON5c}) the overline has been used to stress that the corresponding quantities (unlike the ones of the previous section) are evaluated in the longitudinal gauge; 
$\sigma_{\nu}$ and ${\mathcal F}_{\nu\,3}$ do not have 
a the overline since they are both invariant under infinitesimal coordinate transformations; the higher multipoles (i.e. $\ell\geq 3$) are also gauge-invariant and obey the same equation already reported in section \ref{sec2} (see, in particular, Eq. (\ref{SYN6c})). 

Following the same procedure already outlined in the synchronous coordinate system, we take the conformal time derivative of both sides of Eq. (\ref{LON5c}); we thus obtain the analog of Eq. (\ref{SYN7}):
\begin{equation}
\sigma_{\nu}^{\prime\prime} = \frac{k^2}{15} \overline{\delta}_{\nu} + \frac{4}{15} k^2 \phi - \frac{11}{21}  k^2 \sigma_{\nu},
\label{LON6}
\end{equation}
where the neutrino hierarchy has been truncated, for illustration, to the octupole (notice, however, that ${\mathcal F}_{\nu\, 3}' \neq 0$).
From Eq. (\ref{LON6}) it also follows that:
\begin{equation}
\sigma_{\nu}^{\prime\prime\prime} + \frac{6}{7} k^2 \sigma_{\nu}' = \frac{4 k^2}{15} (\phi -\psi)^{\prime} + \frac{8}{15} k^2 \psi'.
\label{LON7}
\end{equation}
In Eq. (\ref{LON7}) the term $k^2(\phi - \psi)^{\prime}$ can be replaced by taking the derivative of both sides 
of Eq. (\ref{LON5}); the other term appearing at  the right hand side of Eq. (\ref{LON7}) is instead replaced by taking the 
derivative of Eq. (\ref{LON2}) and by inserting, in the obtained expression, the decoupled equation for ${\mathcal R}$, i.e. Eq. (\ref{SYN5}).
The result in terms of $k^2 \psi'$ becomes:
\begin{eqnarray}
k^2 \psi' &=& \frac{1}{c_{\mathrm{st}}^2} \biggl({\mathcal H} - \frac{{\mathcal H}'}{{\mathcal H}}\biggl) (  \frac{{\mathcal H}'}{{\mathcal H}} - 2{\mathcal H} ) ({\mathcal R}' - \Sigma_{{\mathcal R}})  
\nonumber\\
&+& 6 {\mathcal H}^3 \Omega_{R} R_{\nu} \biggl( \sigma_{\nu} + \frac{R_{\gamma}}{R_{\nu}} \sigma_{B} \biggr) - \biggl({\mathcal H} - \frac{{\mathcal H}'}{{\mathcal H}}\biggr) k^2 {\mathcal R}.
\label{LON8}
\end{eqnarray}
Inserting Eq. (\ref{LON8}) inside Eq. (\ref{LON7}) and eliminating $k^2(\phi - \psi)^{\prime}$ with the derivative of Eq. (\ref{LON5}) we obtain the 
equation already reported in Eqs. (\ref{SYN10})  and coinciding, after some algebra, with the result mentioned in  Eq. (\ref{Rseventh}) if the 
background relations discussed in Eq. (\ref{TWOA}) are used.

We conclude this section by recalling that another practical gauge where the derivation can be swiftly performed is the uniform 
curvature gauge \cite{h0,h1,h2} (see also \cite{mm2} second paper). For reasons of space we shall not pursue this discussion here even if 
the general procedure will follow the general lines already discussed in the longitudinal and in the synchronous gauges.

\renewcommand{\theequation}{4.\arabic{equation}}
\setcounter{equation}{0}
\section{Symmetries of the coupled system}
\label{sec4}
The gauge-dependent derivations presented in sections \ref{sec2} and \ref{sec3} led to the gauge-invariant result 
 anticipated in Eqs. (\ref{Rfirst}) and (\ref{Rseventh}). The same conclusion could be reached through a formalism that is explicitly gauge-invariant at every step.
The second issue addressed hereunder involves the possibility of replacing ${\mathcal R}$ with different gauge-invariant variables related to it via the Hamiltonian constraint. It will be argued  that this strategy is inconvenient. Towards the end of the section we shall focus on the dynamical symmetries of the gauge-invariant system and suggest possible generalizations.

\subsection{Gauge-invariant derivations}
The essentials of a fully gauge-invariant derivation are immediate in the light of the results of section \ref{sec3}. It suffices to recall, in fact,  that in the general situation (and in Fourier space),  the scalar fluctuations of the four-dimensional metric are parametrized by four 
different functions: 
\begin{eqnarray}
 \delta_{s} g_{00}(k,\tau) &=& 2 a^2(\tau)\, \phi(k,\tau), \qquad  \delta_{s} g_{0i}(k,\tau) = i  a^2  k_{i}\, \beta(k,\tau),
 \nonumber\\
 \delta_{s} g_{ij}(k,\tau) &=& 2 a^2(\tau) \,[\psi(k,\tau), \delta_{ij} +  k_{i}\, k_{j} \,\alpha(k,\tau)],
\label{UC0} 
\end{eqnarray}
where, as already mentioned, $\delta_{s}$ denotes the scalar mode of the corresponding perturbed entry. 
For infinitesimal coordinate shifts  $\tau \to \widetilde{\tau} = \tau + \epsilon_{0}$ and 
$ {x}^{i} \to \widetilde{x}^{i} = x^{i} + \partial^{i}\epsilon$ the functions $\phi(k,\tau)$, $\beta(k,\tau)$, 
$\psi(k,\tau)$ and $\alpha(k,\tau)$ introduced in Eq. (\ref{UC0}) transform as:
\begin{eqnarray}
&& \phi \to \widetilde{\phi} = \phi - {\mathcal H} \,\epsilon_0 - \epsilon_{0}' ,\qquad \psi \to \widetilde{\psi} = \psi + {\mathcal H}\, \epsilon_{0},
\label{phipsi}\\
&& \beta \to \widetilde{\beta} = \beta +\epsilon_{0} - \epsilon',\qquad \alpha \to \widetilde{\alpha} = \alpha - \epsilon.
\label{EB}
\end{eqnarray}

Equations (\ref{SYN5}) and (\ref{TWOA}) (or, which is the same, Eqs. (\ref{Rfirst}) and (\ref{Rseventh})) 
are derivable in a formalism that is  gauge-invariant at every step by appealing to the properties of the two 
Bardeen potentials \cite{bard1} which are constructed from the fluctuations of Eq. (\ref{UC0}) and are given by:
\begin{equation}
 \Phi= \phi + {\mathcal H} ( \beta - \alpha') + ( \beta - \alpha')', \qquad \Psi = \psi - {\mathcal H} (\beta - \alpha').
\label{GI}
\end{equation}
In the longitudinal gauge $\Psi \equiv \psi$ and $\Phi \equiv \phi$. This means that the equations obeyed by the Bardeen potentials have, by definition, the 
same form of the equations written in the longitudinal gauge. Therefore the wanted gauge-invariant derivation 
will be exactly the one already presented in the case of the conformally Newtonian gauge (see section \ref{sec3}) with the 
proviso that $\phi \to \Phi$ and $\psi \to \Psi$.

\subsection{Different gauge-invariant variables}

The second point we ought to discuss in the present section has to do with the possibility of finding other gauge-invariant variables 
that could play effectively  the role of ${\mathcal R}$. 
In terms of $\Phi$ and $\Psi$ and in real space the Hamiltonian constraint reads
\begin{equation}
\nabla^2 \Psi - 3 {\mathcal H} ({\mathcal H}\Phi + \Psi') = 4 \pi G a^2 (\delta_{s} \rho_{t} + \delta_{s} \rho_{B} +\delta_{s}\rho_{E} ),
\label{GI1}
\end{equation}
where we restored, for the sake of generality, the presence of the electric variables. Let us now recall that 
the curvature perturbations on comoving orthogonal hypersurfaces (i.e. ${\mathcal R}$) and the total density contrast 
on uniform curvature hypersurfaces (conventionally denoted by $\zeta$) are simply related as 
\begin{equation}
 \zeta = {\mathcal R} + \frac{\nabla^2 \Psi}{ 12 \pi G a^2 ( p_{t} + \rho_{t})}.
\label{GI2}
\end{equation} 
Equation (\ref{GI2}) can be derived from Eq. (\ref{GI1}) by recalling 
the gauge-invariant expression of ${\mathcal R}$ and $\zeta$ (see e. g. \cite{mgbook})
\begin{equation}
{\mathcal R} = - \Psi - \frac{{\mathcal H}  ( {\mathcal H} \Phi + \Psi')}{{\mathcal H}^2 - {\mathcal H}'},\qquad 
\zeta = -   \Psi - {\cal H} \frac{\delta_{s}\rho_{t} + \delta_{s}\rho_{B}+ \delta_{s}\rho_{E}}{\rho_{\mathrm{t}}'}. 
\label{GI3}
\end{equation}
where, as in Eq. (\ref{GI1}), $\delta_{s}\rho_{t}$ denotes the gauge-invariant fluctuation of the total energy density. 
It can be shown, by direct calculation, that the exact equation obeyed by $\zeta$ is much more 
involved than the one obeyed by ${\mathcal R}$ even if the two equations coincide in the $k\to 0$ limit. The equation for $\zeta$ can still be written 
in a decoupled form but it contains also terms proportional to $1/[k^2 + {\mathcal H}^2 u(\tau)]$ where $u(\tau)$ generically represents 
a time dependent  function parametrizing the dominant contribution in the $k \to 0$ limit. This aspect can be appreciated by noticing that:
\begin{equation} 
{\mathcal R}' = \Sigma_{{\mathcal R}} - \frac{a^2 \nabla^2 \Psi}{4\pi G {\mathcal H} z_{t}^2}.
\label{GI4}
\end{equation}
If we now use Eq. (\ref{GI2}) to eliminate $\nabla^2 \Psi$, we shall have that $\zeta$ is fully determined 
by ${\mathcal R}$, ${\mathcal R}'$ and $\Sigma_{{\mathcal R}}$ according to the 
relation 
\begin{equation}
\zeta = {\mathcal R} + \frac{\Sigma_{\mathcal R} - {\mathcal R}'}{3 {\mathcal H} c_{\mathrm{st}}^2}.
\label{GI5}
\end{equation}
Equation (\ref{GI5}) quantifies the difference between $\zeta$ and ${\mathcal R}$. This difference depends 
on $\Sigma_{{\mathcal R}}$ and ${\mathcal R}'$. It also suggests that $\zeta$ and ${\mathcal R}$ are 
not fully equivalent when $\Sigma_{{\mathcal R}}\neq 0$. If we ought to avoid 
approximations the best strategy is to compute ${\mathcal R}$ by solving Eq. (\ref{Rfirst}); derive then ${\mathcal R}'$ and finally 
obtain $\zeta$ from Eq. (\ref{GI5}). 

In summary we can say that there exist other variables that are equivalent 
to ${\mathcal R}$ in the large-scale limit. However the use of these variables is inconvenient
since they obey more cumbersome equations. The best strategy is to compute ${\mathcal R}$ and then derive 
all the other gauge-invariant variables.

\subsection{Symmetries and generalizations of the gauge-invariant system}

Having discussed all the technical aspects of the derivation of our system of equations
we shall now discuss tits dynamical symmetries.
Let us therefore consider the case of the vanilla $\Lambda$CDM paradigm and 
write Eqs. (\ref{Rfirst}) and (\ref{Rseventh}) as:
\begin{eqnarray}
&& {\mathcal R}'' + 2 \frac{z_{t}'}{z_{t}} {\mathcal R}'  + k^2 c_{\mathrm{st}}^2 {\mathcal R} = \Sigma_{\mathcal R}' + 2 \frac{z_{t}'}{z_{t}} \Sigma_{\mathcal R}
+ \frac{ 4 c_{\mathrm{st}}^2 }{(1 + w_{t}) } R_{\nu} {\mathcal H}^2  \Omega_{R} \biggl( \sigma_{\nu} + \frac{R_{\gamma}}{R_{\nu}} \sigma_{B}\biggr),
\label{ONE}\\
&& \sigma_{\nu}''' + \frac{8}{5} {\mathcal H}^2 R_{\nu} \Omega_{R} \sigma_{\nu}' + \frac{6}{7} k^2 \sigma_{\nu}' - \frac{32}{5}  {\mathcal H}^3 R_{\nu} \Omega_{R} 
\biggl( \sigma_{\nu} + \frac{R_{\gamma}}{R_{\nu}} \sigma_{B}\biggr)
\nonumber\\
&&= \frac{8}{15 c_{\mathrm{st}}^2} \biggl( {\mathcal H} - \frac{{\mathcal H}'}{{\mathcal H}}\biggr) \biggl(\frac{{\mathcal H}'}{{\mathcal H}} - 2 {\mathcal H}\biggr) ({\mathcal R}' - \Sigma_{{\mathcal R}} ) + \frac{8}{15} \biggl(\frac{{\mathcal H}'}{{\mathcal H}} - {\mathcal H}\biggr) \, k^2 {\mathcal R};
\label{TWO}
\end{eqnarray}
Note that Eq. (\ref{TWO}) coincides with Eq. (\ref{SYN10}) which has been rewritten here for convenience.
For $k \to 0$ Eqs. (\ref{ONE}) and (\ref{TWO}) are left unchanged if 
\begin{eqnarray}
\Sigma_{{\mathcal R}}(k,\tau) &\to& \widetilde{\Sigma}_{{\mathcal R}}(k,\tau) = \Sigma_{{\mathcal R}}(k,\tau) + \frac{\partial {\mathcal A}_{1}}{\partial\tau},
\label{THREE}\\
{\mathcal R}(k,\tau) &\to& \widetilde{{\mathcal R}}(k,\tau) =  {\mathcal R}(k,\tau) + {\mathcal A}_{1}(k,\tau).
\label{FOUR}
\end{eqnarray}

In the $\Lambda$CDM case it is also easy to obtain 
the explicit equation for the total anisotropic stress by defining the shifted 
variable $\sigma_{t}$ and by recalling that $\sigma_{B}'=0$  whenever 
the evolution of the magnetic field is frozen-in. Let us then present a general expression for the evolution of $\sigma_{t}$ recalling that, in our case,
$\Pi_{t} = (p_{t}+ \rho_{t}) \sigma_{t} = ( p_{\nu} + \rho_{\nu}) \sigma_{\nu} + (p_{\gamma} + \rho_{\gamma}) \sigma_{B}$.
The equation for $\sigma_{t}$ becomes, in Fourier space, 
\begin{equation}
\sigma_{t}^{\prime\prime\prime} + {\mathcal M}({\mathcal H}, c_{\mathrm{st}}^2) \sigma_{t}^{\prime\prime} + {\mathcal N}({\mathcal H}, c_{\mathrm{st}}^2, k) \sigma_{t}' + {\mathcal Q}({\mathcal H}, c_{\mathrm{st}}^2, k)  \sigma_{t} = {\mathcal P}({\mathcal H}, c_{\mathrm{st}}^2, k) 
\label{FIVE}
\end{equation}
where 
\begin{eqnarray}
{\mathcal M}({\mathcal H}, c_{\mathrm{st}}^2) &=& 3 {\mathcal H} ( 1 - 3 c_{\mathrm{t}}^2),
\nonumber\\
 {\mathcal N}({\mathcal H}, c_{\mathrm{st}}^2, k) &=& \frac{8}{5} {\mathcal H}^2 R_{\nu}  \Omega_{R} 
 + 3 ( 1 - 3 c_{\mathrm{st}}^2) ({\mathcal H}' + {\mathcal H}^2) +\frac{6}{7} k^2,
 \nonumber\\
  {\mathcal Q}({\mathcal H}, c_{\mathrm{st}}^2, k) &=&\frac{8}{5} {\mathcal H}^3 R_{\nu} \Omega_{R} (1 - 3 c_{\mathrm{st}}^2) 
  \nonumber\\
  &+& 3 {\mathcal H} ( 1 - 3 c_{\mathrm{st}}^2) \biggl( 3 {\mathcal H}' + {\mathcal H}^2 +\frac{{\mathcal H}''}{{\mathcal H}} \biggr) - \frac{96}{5} {\mathcal H}^3 
  c_{\mathrm{st}}^2,
  \nonumber\\
   {\mathcal P}({\mathcal H}, c_{\mathrm{st}}^2, k) &=& \frac{4 z_{t}^2  \, c_{\mathrm{st}}^2}{5 \overline{M}_{\mathrm{P}}^2} R_{\nu} \biggl[ \biggl(\frac{{\mathcal H}}{a^2}\biggr)^{\prime} 
   ( {\mathcal R}^{\prime} - \Sigma_{{\mathcal R}}) - k^2  c_{\mathrm{st}}^2 \biggl(\frac{{\mathcal H}}{a^2}\biggr)\biggr].
\end{eqnarray}
In the case of an exact radiation background we have that $ 3 c_{\mathrm{st}}^2 \to 1$ and 
\begin{eqnarray}
{\mathcal M}({\mathcal H}, c_{\mathrm{st}}^2) &\to& 0,
\nonumber\\
 {\mathcal N}({\mathcal H}, c_{\mathrm{st}}^2, k) &\to& \frac{8}{5} {\mathcal H}^2 R_{\nu}  \Omega_{R} +\frac{6}{7} k^2,
\nonumber\\
 {\mathcal Q}({\mathcal H}, c_{\mathrm{st}}^2, k) &\to& - \frac{32}{5} {\mathcal H}^3,
 \nonumber\\
  {\mathcal P}({\mathcal H}, c_{\mathrm{st}}^2, k) &\to& \frac{4 z_{t}^2}{15 \overline{M}_{\mathrm{P}}^2} R_{\nu} \biggl[ \biggl(\frac{{\mathcal H}}{a^2}\biggr)^{\prime} 
   ( {\mathcal R}^{\prime} - \Sigma_{{\mathcal R}}) - k^2  c_{\mathrm{st}}^2 \biggl(\frac{{\mathcal H}}{a^2}\biggr)\biggr].
\end{eqnarray}
Further generalizations 
of our set of equations are possible but we shall not indulge in these details. We shall now focus on the physical 
properties of the analytical and numerical solutions of the system.

\renewcommand{\theequation}{5.\arabic{equation}}
\setcounter{equation}{0}
\section{Analytic solutions}
\label{sec5}
\subsection{Preliminaries}
Deep in the radiation-dominated epoch and for typical wavelengths larger than the Hubble radius (i.e. $k \tau \ll 1$)  
the total anisotropic stress should obey the conditions 
\begin{equation}
\sigma_{t}(k,\tau) \ll 1 ,\qquad \sigma_{t}'(k,\tau) \ll 1, \qquad  \sigma_{t}''(k,\tau) \ll 1.
\label{ANN1}
\end{equation}
The conditions of Eq. (\ref{ANN1}) can be guessed from the known properties of the standard adiabatic solution \cite{admode} (see also 
\cite{weinbook,mgbook}). In the vanilla $\Lambda$CDM paradigm the dimensionless anisotropic stress is 
$\sigma_{t} = R_{\nu} \sigma_{\nu}$. Recalling the symmetries  of Eqs. (\ref{THREE})--(\ref{FOUR}), 
in the magnetized $\Lambda$CDM paradigm (sometimes dubbed m$\Lambda$CDM) $\sigma_{t} =  R_{\nu} \sigma_{\nu} + R_{\gamma}\sigma_{B}$.

The initial conditions for ${\mathcal R}(k,\tau)$ during the radiation-dominated phase can be bootstrapped from the inflationary solution: 
\begin{equation}
{\mathcal R}(k,\tau) \simeq {\mathcal R}_{*}(k) + \frac{1 + \eta + \epsilon}{(3 - \epsilon)} \biggl(\frac{z_{*}}{z_{t}} \biggr)^2 
\biggl(\frac{H_{*} a_{*}}{H a}\biggr) {\mathcal R}_{*},
\label{ANN2}
\end{equation}
where the star denotes the time at which the given wavelength crossed the Hubble radius during inflation; in Eq. (\ref{ANN2}) $\epsilon= - \dot{H}/H^2$ and $\eta = \ddot{\varphi}/(H\dot{\varphi})$ are the two standard slow-roll parameters that we take to be constant, for simplicity. 
Assuming a sudden reheating approximation taking place at a typical conformal time-scale 
$\tau_{r}$ the solution for the evolution of the curvature perturbations with wavelength larger 
than the Hubble radius can be written, after some simple algebra: 
\begin{equation}
{\mathcal R}(k,\tau) = {\mathcal R}_{*}(k) + \biggl(\frac{a_{r}}{a} \biggr) \frac{(1 + \eta + \epsilon)( 3 + \eta + \epsilon)}{(3 - \epsilon)} {\mathcal R}_{*}(k) 
\biggl(\frac{a_{*} \, H_{*}}{a_{r} \, H_{r}}\biggr)^3 
\label{ANN3}
\end{equation}
To higher order in $k\tau$ the decaying mode practically does not affect the solution since it is strongly suppressed and the full solution can
be parametrized as ${\mathcal R}(k,\tau) = {\mathcal R}_{*} j_{0}(c_{\mathrm{st}}\,y)$ where $c_{\mathrm{st}} \simeq 1/\sqrt{3}$ is the sound speed during the radiation-dominated phase and $j_{0}(c_{\mathrm{st}}\,y)$ is the spherical Bessel function of zeroth order.

Equations (\ref{ANN2}) and (\ref{ANN3}) can be directly used as initial conditions of the subsequent evolution but they can 
also be simplified by just positing the constancy of ${\mathcal R}$ at large-scales. Again this is typical of the standard 
adiabatic solution incorporated in the $\Lambda$CDM paradigm but the general situation could be different.
During the subsequent radiation epoch $c_{\mathrm{st}} = 1/\sqrt{3}$, and the canonical form of  Eqs. (\ref{ONE})--(\ref{FIVE}) can be directly written in terms  $y = k\tau$:
\begin{eqnarray}
&& \frac{d^2 {\mathcal R}}{d y^2} + \frac{2}{y} \frac{d {\mathcal R}}{d y} + \frac{{\mathcal R}}{3} = \frac{1}{y} \biggl( \frac{d \sigma_{t}}{d y} + \frac{2}{y} \sigma_{t} \biggr),
\label{ANN4}\\
&& \frac{d^3 \sigma_{t}}{d y^3} + \biggl(\frac{6}{7} + \frac{8 R_{\nu}}{5 y^2} \biggr)\frac{d \sigma_{t}}{d y} - 16 \frac{R_{\nu}}{y^3}\sigma_{t}
 + \frac{16 R_{\nu}}{5 y^2 } \biggl( 3 \frac{d {\mathcal R}}{d y} +  \frac{y {\mathcal R}}{3}  \biggr) =0.
\label{ANN5}
\end{eqnarray}
The variable $y$ during the radiation epoch obeys the exact chain of equalities $y = k/{\mathcal H} = k/(a H)$ since, during 
radiation, ${\mathcal H} = 1/\tau$. This means that $y$ can be viewed as the ratio between the particle horizon and 
the physical wavelength of the fluctuation. Equations (\ref{ANN4})--(\ref{ANN5}) imply that the system depends on a single 
scaling variable.

\subsection{Solutions in the radiation epoch}

We look preliminarily for an approximate solution 
 of   Eqs.  (\ref{ANN4}) and (\ref{ANN5})  in the limit $y < 1$ and with the initial conditions of Eq.  (\ref{ANN1}):
\begin{eqnarray}
&& \sigma_{t}(y) = {\mathcal A} y^{\gamma}  + {\mathcal O}(y^{2 + \gamma}), \qquad \gamma >0,
\nonumber\\
&& {\mathcal R}(y) = {\mathcal R}_{*} + {\mathcal B} y^{\delta} + {\mathcal O}(y^{2 + \delta}), \qquad \delta> 0.
\label{ANN10}
\end{eqnarray}
Equations (\ref{ANN4}) and (\ref{ANN5}) imply that the parameters of Eq. (\ref{ANN10}) must obey the following relation:
\begin{equation}
{\mathcal A} = \frac{18 {\mathcal B} + {\mathcal R}_{*}}{12}, \qquad \delta = \gamma = 2.
\label{ANN11}
\end{equation}
The gauge-invariant solution of Eqs. (\ref{ANN10})--(\ref{ANN11}) can be translated in any specific coordinate system and, in particular, 
 in the conformally Newtonian gauge:
\begin{equation}
\frac{d {\mathcal R}}{d y} = \frac{\sigma_{t}}{y}  + \frac{y \psi}{6}, \qquad {\mathcal R} = - \psi - \frac{\phi}{2}.
\label{ANN11a}
\end{equation}
Eq. (\ref{ANN11}) implies ${\mathcal A} = (\psi - \phi)/6 $ where $\psi$ and $\phi$ are both constant to leading order: this 
is the standard result valid in the longitudinal gauge (see, e.g. \cite{admode,pee} and also \cite{mm1,mm2}). The same strategy can be used in the  synchronous coordinate system. The full gauge-invariant solution for the magnetized adiabatic mode can be written explicitly as \cite{mm1,mm2}:
\begin{eqnarray}
{\mathcal R}(y) &=& {\mathcal R}_{*} + {\mathcal B} y^2 + {\mathcal O}(y^4),
\nonumber\\
\sigma_{\nu} ( y) &=& - \frac{R_{\gamma}}{R_{\nu}} \sigma_{B} +  \frac{18 {\mathcal B} + {\mathcal R}_{*}}{12} y^2 + {\mathcal O}(y^4).
\label{ANN11b}
\end{eqnarray}
In the limit $\sigma_{B}\to 0$, the solution of Eq. (\ref{ANN11b}) reproduces the standard adiabatic mode.

From the previous solution we can also derive the solution for $\zeta$; in fact Eq. (\ref{GI5}) implies
\begin{equation}
\zeta(y) = {\mathcal R}(y) + \sigma_{t}(y) - y \frac{d {\mathcal R}}{d y}  \equiv {\mathcal R}_{*}+ \frac{1}{2}\biggl( {\mathcal B} + \frac{{\mathcal R}_{*}}{6}\biggr) y^2 + {\mathcal O}(y^4),
\label{ANN13}
\end{equation}
where the second equality follows after inserting the solution of Eq. (\ref{ANN11b}) into the first relation of Eq. (\ref{ANN13}).
If we take now the difference between $\zeta$ and ${\mathcal R}$ we correctly obtain 
the Hamiltonian constraint stipulating that 
\begin{equation}
\zeta - {\mathcal R} = \frac{1}{2} \biggl( \frac{{\mathcal R}_{*}}{6} - {\mathcal B}\biggr) y^2 \equiv - \frac{\psi_{*}}{6} y^2.
\label{ANN14}
\end{equation}
Since ${\mathcal R}(y)$ and $\sigma_{\nu}(y)$ are both gauge-invariant the explicit values of ${\mathcal A}$ and ${\mathcal B}$ are 
also gauge-invariant and can be determined in any gauge by demanding that the Hamiltonian 
and the momentum constraints are satisfied. Skipping the technical details the results of this analysis are:
\begin{eqnarray}
{\mathcal A}({\mathcal R}_{*},\, \Omega_{B},\,\sigma_{B}) &=& - \frac{2 R_{\nu} {\mathcal R}_{*}}{3 (15 + 4 R_{\nu})} - \frac{R_{\gamma} \,R_{\nu} \, ( R_{\nu} \Omega_{B} - 4 \sigma_{B})}{ 2 R_{\nu} ( 15 + 4 R_{\nu})}, 
\label{Aex}\\
{\mathcal B}({\mathcal R}_{*},\, \Omega_{B},\,\sigma_{B}) &=& - \frac{(5 + 4 R_{\nu}) {\mathcal R}_{*}}{6 (15 + 4 R_{\nu})} - \frac{R_{\gamma} ( R_{\nu} \Omega_{B} - 4 \sigma_{B})}{3( 15 + 4 R_{\nu})}.
\label{Bex}
\end{eqnarray}

The initial conditions during radiation can be fixed by demanding that the only source of anisotropic stress is provided by the magnetic fields. In this case the equation to integrate is simply given by:
\begin{equation}
\frac{d^{2} {\mathcal R}}{d y^2} + \frac{2}{y} \frac{d R}{d y} + c_{\mathrm{st}}^2 {\mathcal R} = \frac{ 2 R_{\gamma} \sigma_{B}}{y^2}.
\label{ANN17}
\end{equation}
Defining $q(y) = y {\mathcal R}(y)$ we get the following  equation
\begin{equation}
\frac{d^2 q}{d y^2} + c_{\mathrm{st}}^2 q = \frac{2 R_{\gamma} \sigma_{B}}{y},
\label{ANN18}
\end{equation}
whose general solution can be written as:
\begin{equation}
q(y) = C_{1} \cos{(c_{\mathrm{st}} y)} + C_{2} \sin{(c_{\mathrm{st}} y)} + \frac{2 R_{\gamma} \sigma_{B} }{c_{\mathrm{st}}} \int_{y_{i}}^{y} \, \frac{\sin{[c_{\mathrm{st}} (y - \xi)]}}{\xi} \, d\xi.
\label{ANN19}
\end{equation}
In terms of ${\mathcal R}(y)$ the solution of Eq. (\ref{ANN19}) with the large-scale boundary conditions 
provided by the adiabatic mode can be written as 
\begin{eqnarray}
{\mathcal R}(y) &=& {\mathcal R}_{*} \frac{\sin{(c_{\mathrm{st}} y)}}{c_{\mathrm{st}} y}  
\nonumber\\
&+& 2 R_{\gamma} \sigma_{B} \biggl[ \biggl( \mathrm{Ci}(c_{\mathrm{st}} y) -   \mathrm{Ci}(c_{\mathrm{st}} y_{i}) \biggr) \frac{\sin{(c_{\mathrm{st}} y)}}{c_{\mathrm{st}} y} 
\nonumber\\
&+& \biggl( \mathrm{Si}(c_{\mathrm{st}} y_{i}) -   \mathrm{Si}(c_{\mathrm{st}} y) \biggr) \frac{\cos{(c_{\mathrm{st}} y)}}{c_{\mathrm{st}} y} \biggr],
\label{ANN20}
\end{eqnarray}
where 
\begin{equation}
\mathrm{Ci}(z) = - \int_{z}^{\infty} \frac{\cos{t}}{t} \, dt, \qquad  \mathrm{Si}(z) = \int_{0}^{z} \frac{\sin{t}}{t}\, dt.
\end{equation}
The previous expression can be expanded in powers of $y$ and $y_{i}$ with the result that:
\begin{equation}
{\mathcal R}(y) = {\mathcal R}_{*} + 2 R_{\gamma} \sigma_{B} [\ln{(y/y_{i})} -1] + 2 R_{\gamma} \sigma_{B} y_{i}/y + {\mathcal O}(y^2) + {\mathcal O}(y^2 y_{i}) + {\mathcal O}(y_{i}^2) 
\label{ANN21}
\end{equation}
where we recall that $y \geq y_{i}$. As we shall see in section \ref{sec6} the purely magnetic initial conditions can be 
studied from the more general perspective of a consistent formulation of the Cauchy problem for the coupled 
evolution of the quasinormal mode and of the total anisotropic stress.

\subsection{Matter-radiation transition}

The equations for the curvature perturbations and for the shifted anisotropic stress can be written 
in the $\alpha$ paramerization where $\alpha = a/a_{eq}$ is the normalized 
scale factor\footnote{The notation employed here does not conflict with the one employed in Eq. (\ref{UC0}) since the two variables are never used 
together in the same context or even in neighbouring discussions.}. The exact solution of Friedmann equations (\ref{FL1})--(\ref{FL3}) across the transition implies that 
$\alpha = (\tau/\tau_{1})^2 + 2 (\tau/\tau_{1})$; $\tau_{1} = (\sqrt{2} +1) \tau_{eq}$ and 
$\tau_{eq}$ denotes the time of matter-radiation equality.  In the $\alpha$ parametrization we have:
\begin{eqnarray}
&& \tau_{1} \Sigma_{\mathcal R}(\alpha) = \frac{8 \sqrt{\alpha + 1}}{\alpha ( 3 \alpha + 4)} \biggl[ \sigma_{t} 
- \frac{3 \alpha \Omega_{B} R_{\gamma}}{4 ( 3\alpha + 4)} \biggr],
\nonumber\\
&& z_{t}^2(\alpha) = \frac{3}{4} \,\overline{M}_{P}^2 \,a_{eq}^2 \, \frac{\alpha^2 ( 3 \alpha + 4)^2}{\alpha + 1}.
\label{RS}
\end{eqnarray}
The evolution equation of the quasinormal mode
can be written as:
\begin{equation}
\frac{\partial^2 {\mathcal R}}{\partial \alpha^2} + {\mathcal Q}_{1}(\alpha) \frac{\partial {\mathcal R}}{\partial \alpha} 
+ {\mathcal Q}_{2}(\alpha) \kappa^2   {\mathcal R}  =   \biggl[ {\mathcal Q}_{3}(\alpha) \sigma_{t} + {\mathcal Q}_{4}(\alpha)\frac{\partial \sigma_{t}}{\partial \alpha} \biggr]- {\mathcal Q}_{5}(\alpha) R_{\gamma} \Omega_{B},
\label{Ralpha}
\end{equation}
where the five background dependent functions of Eq. (\ref{Ralpha}) are:
\begin{eqnarray}
 {\mathcal Q}_{1}(\alpha) &=& \frac{ 21 \alpha^2 + 36 \alpha + 16}{2 \alpha (3\alpha + 4) (\alpha+1)}, \qquad  {\mathcal Q}_{2}(\alpha) = \frac{1}{3 ( 3\alpha + 4 )(\alpha+ 1)}, 
 \nonumber\\
{\mathcal Q}_{3}(\alpha) &=& \frac{2 ( 9\alpha^2 + 24 \alpha +16)}{\alpha^2 (\alpha + 1) ( 3 \alpha + 4)^2},\qquad {\mathcal Q}_{4}(\alpha) = \frac{4 }{\alpha (3\alpha+4)},
\nonumber\\
 {\mathcal Q}_{5}(\alpha) &=&  \frac{3}{2 \alpha ( \alpha +1 ) ( 3 \alpha + 4)}.
 \label{Ralpha2}
\end{eqnarray}

In the notations of Eq. (\ref{Ralpha}), the equation obeyed by the shifted anisotropic stress is:
\begin{eqnarray}
&&\frac{\partial^3 \sigma_{t}}{\partial \alpha^3} + {\mathcal U}_{1}(\alpha) \frac{\partial^2 
\sigma_{t}}{\partial \alpha^2}  +\biggl[{\mathcal U}_{2}(\alpha) + \kappa^2 {\mathcal U}_{3}(\alpha) \biggr] 
\frac{\partial \sigma_{t}}{\partial \alpha} - {\mathcal U}_{4}(\alpha) \sigma_{t} = - 
\kappa^2 {\mathcal U}_{5}(\alpha) {\mathcal R}  
\nonumber\\
&& - R_{\nu} {\mathcal U}_{6}(\alpha) \frac{\partial {\mathcal R}}{\partial \alpha} 
- {\mathcal U}_{7}(\alpha) R_{\nu} R_{\gamma} \Omega_{B},
\label{Sigmalpha}
\end{eqnarray}
where the background dependent functions are now defined as:
\begin{eqnarray}
&& {\mathcal U}_{1}(\alpha) =\frac{3}{2(\alpha + 1)}, \qquad {\mathcal U}_{2}(\alpha) =\frac{8}{5 \alpha^2(\alpha+1)},
\nonumber\\
&& {\mathcal U}_{3}(\alpha) =\frac{3}{14 (\alpha +1)}, \qquad  {\mathcal U}_{4}(\alpha)= \frac{2(15 \alpha^2 + 54 \alpha +40)}{5 \alpha^3 (\alpha +1)^2},
\nonumber\\
&& {\mathcal U}_{5}(\alpha) = \frac{(3\alpha + 4)}{15 \alpha (\alpha+1)^2}, \qquad {\mathcal U}_{6}(\alpha) = \frac{(3\alpha +4)^2 
(5 \alpha + 6)}{10 \alpha^2 (\alpha +1)^2}, 
\nonumber\\
&& {\mathcal U}_{7}(\alpha) = 
\frac{3 (5 \alpha + 6)}{10\alpha^2 (\alpha +1)^2}.
\label{Sigmalpha2}
\end{eqnarray}
In the limit $\alpha\to 0$, the functions ${\mathcal Q}_{i}(\alpha)$ (with $i$ going from $1$ to $5$) and ${\mathcal U}_{j}(\alpha)$ (with 
$j$ going from $1$ to $7$) can be expanded and the resulting 
system of equations becomes:
\begin{eqnarray}
&& \frac{\partial^2 {\mathcal R}}{\partial \alpha^2} + \frac{2}{\alpha} \frac{\partial {\mathcal R}}{\partial \alpha} 
+  \frac{\kappa^2 }{12}  {\mathcal R}  =   \biggl[\frac{2}{\alpha}\sigma_{t} + \frac{\partial\sigma_{t}}{\partial \alpha} \biggr]-
 \frac{3}{8 \alpha} R_{\gamma} \Omega_{B},
\nonumber\\
&& \frac{\partial^3 \sigma_{t}}{\partial \alpha^3} + \frac{3}{2} \frac{\partial^2 
\sigma_{t}}{\partial \alpha^2}  +\biggl[\frac{8}{5\alpha^2} +\frac{3 \kappa^2}{14}  \biggr] 
\frac{\partial\sigma_{t}}{\partial \alpha} -\frac{16}{\alpha^3} R_{\nu} \sigma_{t} = - 
\frac{4 \kappa^2}{15} R_{\nu} {\mathcal R}  - \frac{48}{5\alpha^2} R_{\nu} \frac{\partial {\mathcal R}}{\partial \alpha} 
-\frac{9}{5\alpha^2}  R_{\nu} R_{\gamma} \Omega_{B}.
\nonumber
\end{eqnarray}
It is clear that these equations coincide with the ones previously derived since $\kappa = k \tau_{1}$ and $\alpha \simeq 2 \,\tau/\tau_{1}$. 
Note that the important terms containing $\Omega_{B}$ are suppressed: if we pass to the variable $ \kappa \alpha = 2 y$ the terms 
containing $\Omega_{B}$ turn out to be suppressed as $\Omega_{M}/\Omega_{R}= \alpha$ (where as usual 
$\Omega_{M} = \rho_{M}/\rho_{t}$) in the limit $\alpha \to 0$. 

\renewcommand{\theequation}{6.\arabic{equation}}
\setcounter{equation}{0}
\section{Setting the large-scale initial conditions}
\label{sec6}
\subsection{Generalities}
So far no attempt has been 
made to formulate the Cauchy problem of large-scale inhomogeneities solely on the basis of the system discussed in the present paper.
To proceed along this direction, it is convenient to change the variable appearing in Eqs. (\ref{ANN4}) and (\ref{ANN5})  from $y$ to $ x= \ln{y}$; with this change of variables the equations become\footnote{In this paper (and following widely used notations) $\ln{}$ is
the natural (i.e. Neperian) logarithm while $\log{}$ denotes the common logarithm (i.e. to base $10$).}:
\begin{eqnarray}
&& \frac{d^{3}\sigma_{\mathrm{t}}}{d x^3} - 3 \frac{d^{2} \sigma_{\mathrm{t}}}{dx^2} + \biggl( \frac{8}{5}  R_{\nu}+ 2 + \frac{6}{7} e^{2 x} \biggr) \frac{d \sigma_{\mathrm{t}}}{d x} 
= 16 R_{\nu} \sigma_{\mathrm{t}} 
 - \frac{48 R_{\nu}}{5} \biggl(\frac{d{\mathcal R}}{d x} + \frac{{\mathcal R}}{9} e^{2 x}\biggr),
\label{ANN6a}\\
&&\frac{d^2 {\mathcal R}}{d x^2} + \frac{d {\mathcal R}}{d x}  + c_{\mathrm{st}}^2 \,e^{ 2 x} {\mathcal R} =  \biggl( \frac{d\sigma_{\mathrm{t}}}{d x} + 2 \sigma_{\mathrm{t}} \biggr).
\label{ANN6}
\end{eqnarray}

Five initial conditions define the space of the Cauchy data of Eqs. (\ref{ANN6a})--(\ref{ANN6}): 
three initial data are related to Eq. (\ref{ANN6a}) and two initial 
data must be specified in connection with Eq. (\ref{ANN6}). We shall use the following shorthand notation\footnote{For the sake of clarity we mention that the prime has been used in the previous sections to denote a derivation with respect to the conformal time coordinate $\tau$ while, only in the present section, the prime will be used to denote a derivation with respect to $x$. There is no contradiction between the two notations since, in both cases, the prime simply denotes a derivation with respect to the argument of each function. We are confident that, with this remark, potential confusions are avoided.}:
\begin{equation}
\sigma_{\mathrm{t}}(x_{i}),\qquad \sigma_{\mathrm{t}}'(x_{i})=\frac{d \sigma_{\mathrm{t}}}{d x}\biggl|_{x= x_{i}} , \qquad \sigma_{\mathrm{t}}''(x_{i})=\frac{d^2 \sigma_{\mathrm{t}}}{d x^2}\biggl|_{x= x_{i}},
\label{INCON0}
\end{equation}
for the total anisotropic stress and its first two derivatives and 
\begin{equation}
 {\mathcal R}(x_{i}), \qquad  {\mathcal R}'(x_{i}) = \frac{d{\mathcal R}}{d x}\biggl|_{x= x_{i}},
\label{INCON1}
\end{equation}
for the curvature perturbations and its first derivative. 
\begin{figure}[!ht]
\centering
\includegraphics[height=6.8cm]{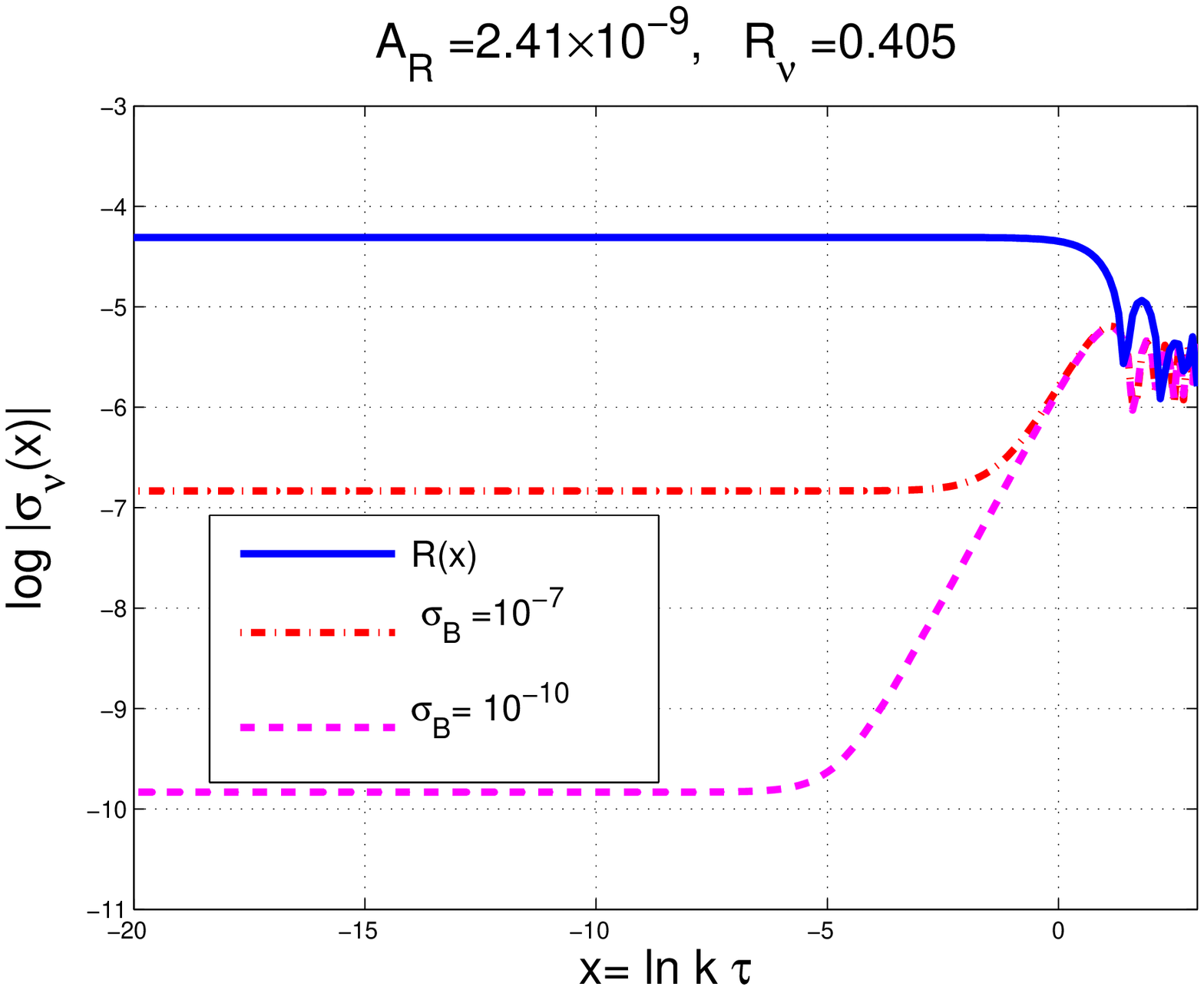}
\includegraphics[height=6.8cm]{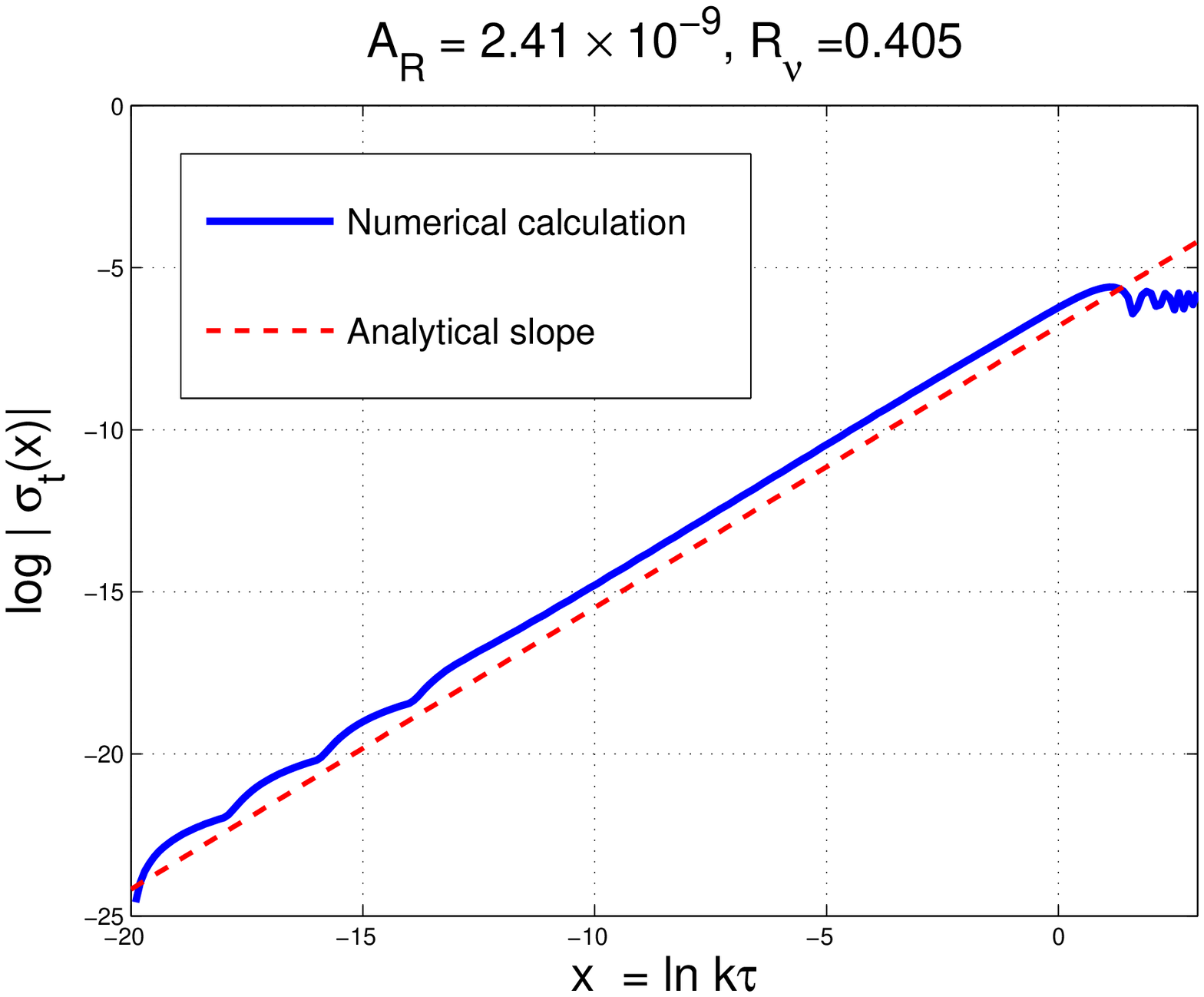}
\caption[a]{The standard adiabatic initial conditions of Eq. (\ref{INCON2}) are illustrated.}
\label{Figure1}      
\end{figure}

The initial data of Eqs. (\ref{INCON0}) and (\ref{INCON1}) are assigned when all the relevant wavelengths are larger than the Hubble radius during the radiation epoch. We define $x_{i}$ as the initial value of $x = \ln{k\tau}$. Notice that $x_{i}$ will typically be larger than one  (in absolute value) and negative. This is because
$k \tau = k/{\mathcal H} \ll 1$ for all the wavelengths larger than the Hubble radius. 
The large-scale power spectra of curvature are customarily assigned at a typical pivot scale  $k_{\mathrm{p}} = 0.002\,\mathrm{Mpc}^{-1}$ (see e.g. \cite{WMAP9a,WMAP9}).
The fiducial values of the cosmological parameters determined, for instance, by the WMAP collaboration \cite{WMAP9}  (see also\footnote{In the 
present analysis we shall use the fiducial values of the WMAP 9yr data alone analyzed in the light of the $\Lambda$CDM paradigm.
The fiducial values of earlier data releases would be equally good for the general purposes of this analysis.}
 \cite{WMAP9a}) imply 
$\tau_{eq} \simeq {\mathcal O}(120)\, \mathrm{Mpc}$ for the equality time already discussed at the end of section \ref{sec5}. 
Thus the maximal value of  $x$ at equality  
be of the order of $\ln{(k_{\mathrm{p}}\, \tau_{\mathrm{eq}})} \sim - 1.4$. The initial value of $x$ is much smaller. For instance, 
if $\tau$ coincides with the time of neutrino decoupling we will have that $k_{\mathrm{p}} \tau_{i}$ can easily be ${\mathcal O}(10^{-9})$ 
so that $x_{i} = \ln{k \tau_{i}} = {\mathcal O}( - 20)$. In the numerical examples 
discussed hereunder we shall always consider a fiducial interval of $x$ ranging from $-20$ to $3$. When $x$ is in the 
interval $ -1.4 < x < 3$  the radiation phase is already replaced by the matter dominated epoch. 
Even of the physical range of $x$ goes from $-20$ to $-1.4$ we shall 
extend this range to positive values of $x$ and artificially assume a slightly longer duration of the radiation phase. This 
is done just to account graphically for the oscillating regime. 
The quantitative error due to the matter-radiation transition is immaterial 
for the present considerations but can become important if we ought to assess the magnetized observables of the Cosmic Microwave Background (see e.g. \cite{mm1,mm2}).

\subsection{The conventional adiabatic paradigm}
As already discussed in section \ref{sec5} in the standard $\Lambda$CDM case the total anisotropic stress 
coincides with the neutrino component and in the language of Eqs. (\ref{INCON0}) and (\ref{INCON1}) the adiabatic initial conditions demand: 
\begin{equation}
 {\mathcal R}(x_{i}) = {\mathcal R}_{*}, \qquad  {\mathcal R}'(x_{i}) =0, \qquad \sigma_{\mathrm{t}}(x_{i}) = \sigma_{\mathrm{t}}'(x_{i}) = \sigma_{\mathrm{t}}''(x_{i})=0.
\label{INCON2}
\end{equation}
The request that a given function equals zero at $x=x_{i}$ means, in physical terms, that its value is 
${\mathcal O}(e^{2 x_{i}})$ for $x_{i} \ll 1$. The numerical integrations must not depend on 
the initial conditions being fixed in terms of (approximate) analytic solution. For this reason we shall rather 
use Eq. (\ref{INCON2}) (and its descendants) and verify that the numerical result is indeed consistent with the analytic estimate.

It is customary, in similar applications, 
to select an exactly scale-invariant spectral amplitude with ${\mathcal R}_{*}= 1$; in this case all the results 
will be effectively given in units of ${\mathcal R}_{*}$. We prefer to set a quantitatively realistic normalization by relating 
the value of ${\mathcal R}_{*}$ to the spectral amplitude as:
\begin{equation}
{\mathcal R}_{*} =  4.9 \times 10^{-5} \biggl(\frac{{\mathcal A}_{{\mathcal R}}}{2.41\times 
10^{-9}}\biggr)^{1/2},
\label{INCON3}
\end{equation}
where the fiducial value of ${\mathcal A}_{{\mathcal R}}$ is compatible with the WMAP releases \cite{WMAP9a} and it coincides, in particular, 
with WMAP 9-years data \cite{WMAP9}. According to Eq. (\ref{INCON3}) the corresponding inflationary curvature scale will be given by:
\begin{equation}
\frac{H}{M_{P}} = \frac{\sqrt{\pi r_{T} {\mathcal A}_{{\mathcal R}}}}{4} =9.72\times 10^{-6} \biggl(\frac{r_{T}}{0.2}\biggr)^{1/2} \, \biggl(\frac{{\mathcal A}_{{\mathcal R}}}{2.41\times 10^{-9}}\biggr)^{1/2},
\label{INCON3a}
\end{equation}
where the fiducial value of the tensor to scalar ratio $r_{T}$ has been estimated though the BICEP2 data \cite{BICEP2}.
Scaling violations of the initial spectrum can be included but shall not be considered here. 

In Fig. \ref{Figure1} the result of the numerical integration is illustrated when the adiabatic initial conditions are set 
according to Eq. (\ref{INCON2}).  In all the graphs of this section we shall report on the vertical axis 
the common logarithm (i.e. to base $10$) of the corresponding quantity; on the horizontal axis we shall 
instead use the natural logarithm of $k\tau$, as implied by the definition of the variable $x$.
In the left panel of Fig. \ref{Figure1} the 
dashed and dot-dashed lines correspond to different values of the magnetic anisotropic stress, as mentioned in the legend.
Always in the left panel of Fig. \ref{Figure1} we illustrate, with the full line, the common logarithm of the curvature perturbation 
${\mathcal R}(x)$. The total anisotropic stress in the absence of the magnetic contribution is reported in the right panel of Fig. \ref{Figure1}. The dashed line is the analytical slope obtained in Eq. (\ref{ANN10}) and (\ref{ANN11}). 

As expected, according to Fig. \ref{Figure1}, the curvature perturbations are always in the linear regime at early times since they are practically constant for typical wavelengths larger than the Hubble radius. This is a peculiar property 
of the adiabatic mode (both with and without magnetic contribution).  The results of Fig. \ref{Figure1} 
confirm that the adiabatic initial data are indeed equivalent, in the present context, to the initial conditions 
of Eq. (\ref{INCON2}). 

\begin{figure}[!ht]
\centering
\includegraphics[height=6.8cm]{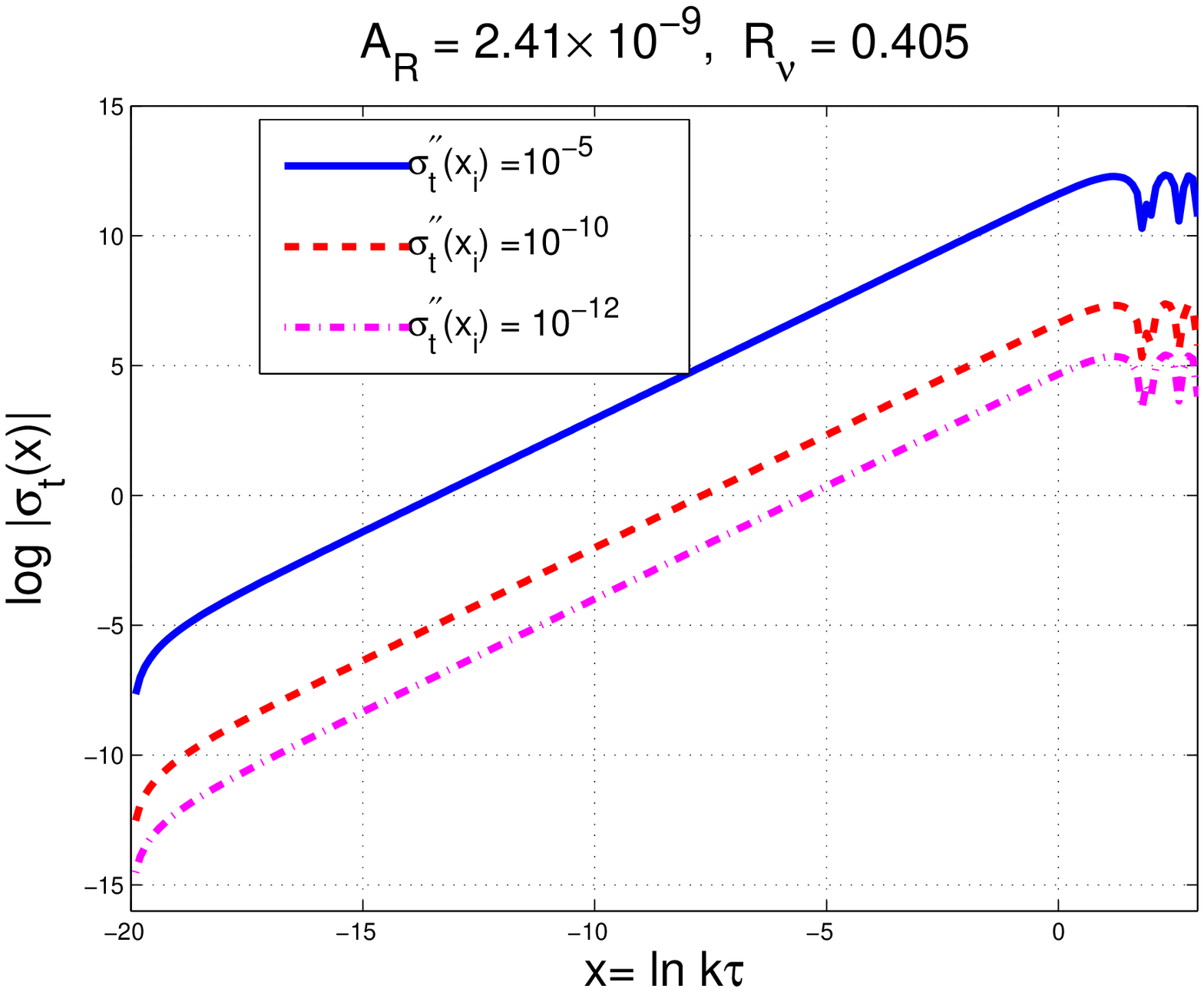}
\includegraphics[height=6.8cm]{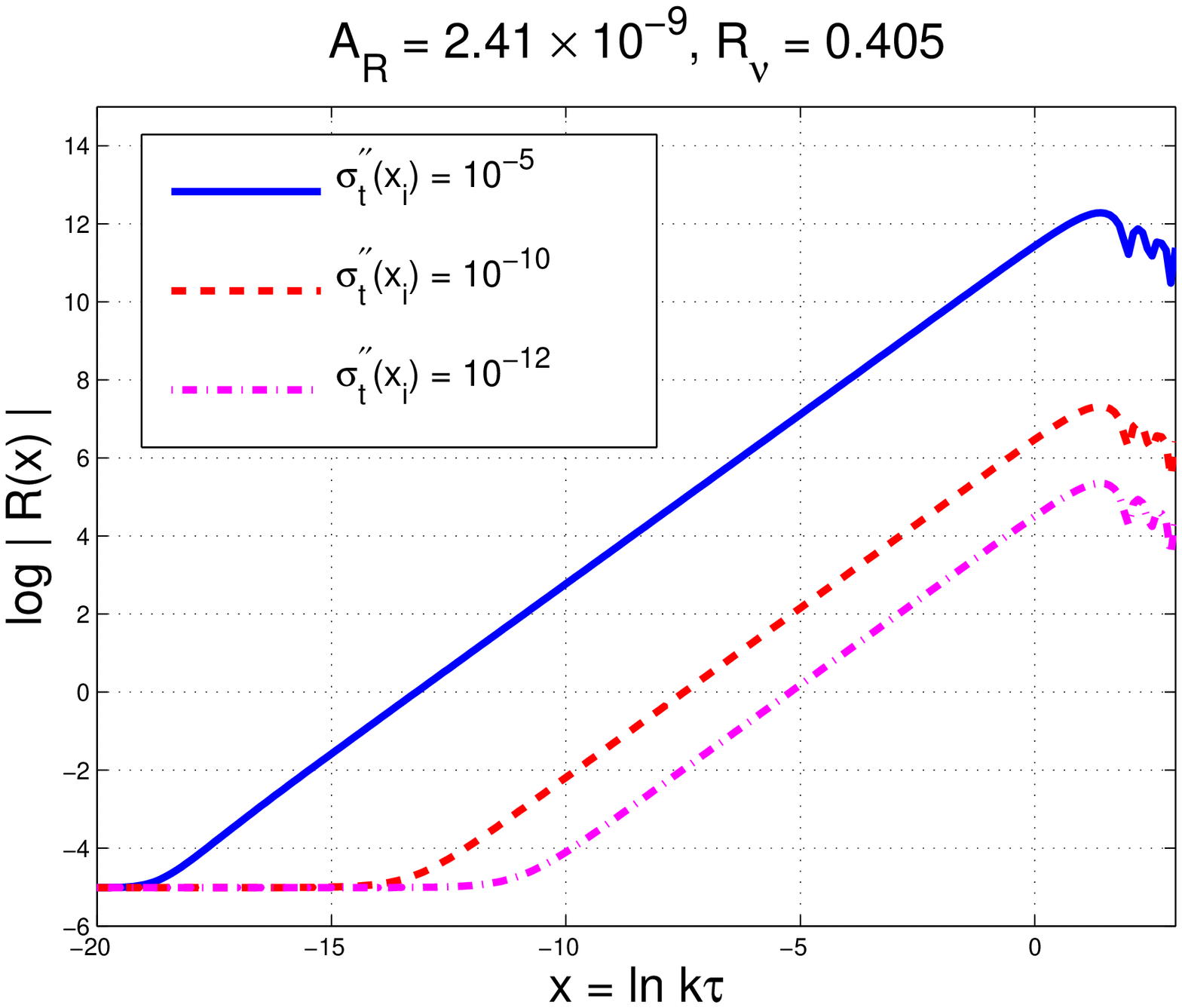}
\caption[a]{Some early departures from the linear approximation are illustrated 
in the case when the second derivative of the anisotropic stress does not vanish.}
\label{Figure2}      
\end{figure}

\subsection{Early departures from the linear approximation}
The constancy of the curvature perturbations for wavelengths 
larger than the Hubble radius is not a generic property of Eqs. (\ref{ANN6a}) and (\ref{ANN6}). It is therefore 
reasonable to perturb slightly the initial data of Eq. (\ref{INCON2}) by allowing, for instance, 
arbitrary derivatives of the total anisotropic stress at $x_{i}$. 

A possible numerical answer to the question of the previous paragraph is illustrated in Fig. \ref{Figure2} within the same notations of Fig. \ref{Figure1}.
In Fig. \ref{Figure2} the initial conditions of Eq. (\ref{INCON2}) have been generalized  
by allowing for a small second derivative of the total anisotropic stress while both $\sigma_{t}(x_{i})$ and $\sigma_{t}'(x_{i})$ are set to zero. 
In the left panel we illustrate 
the total anisotropic stress while in the right panel the curvature 
perturbations are reported. In both panels of Fig. \ref{Figure2}  the full, dashed and dot-dashed lines correspond to three progressively smaller valued of $\sigma_{\mathrm{t}}''(x_{i})$. Taking at face value the result of Fig. \ref{Figure2}, if  $\sigma_{\mathrm{t}}''(x_{i}) = 10^{-10}$ then ${\mathcal R}(x) = {\mathcal O}(1)$ for $x\sim - 7.5$
i.e. for modes that are larger than the Hubble radius during radiation.  Depending on the largeness of $\sigma_{\mathrm{t}}''(x_{i})$ the linear approximation can be violated for progressively earlier times (see, e.g. the full line in the right panel of Fig. \ref{Figure2}). Without 
indulging into details, we mention that various explicit examples similar to the one of Fig. \ref{Figure2} can be devised. 
The numerical evidence can be summarized by saying that, depending on the initial conditions of the total anisotropic stress, the curvature perturbations can suddenly grow. 

\begin{figure}[!ht]
\centering
\includegraphics[height=6.8cm]{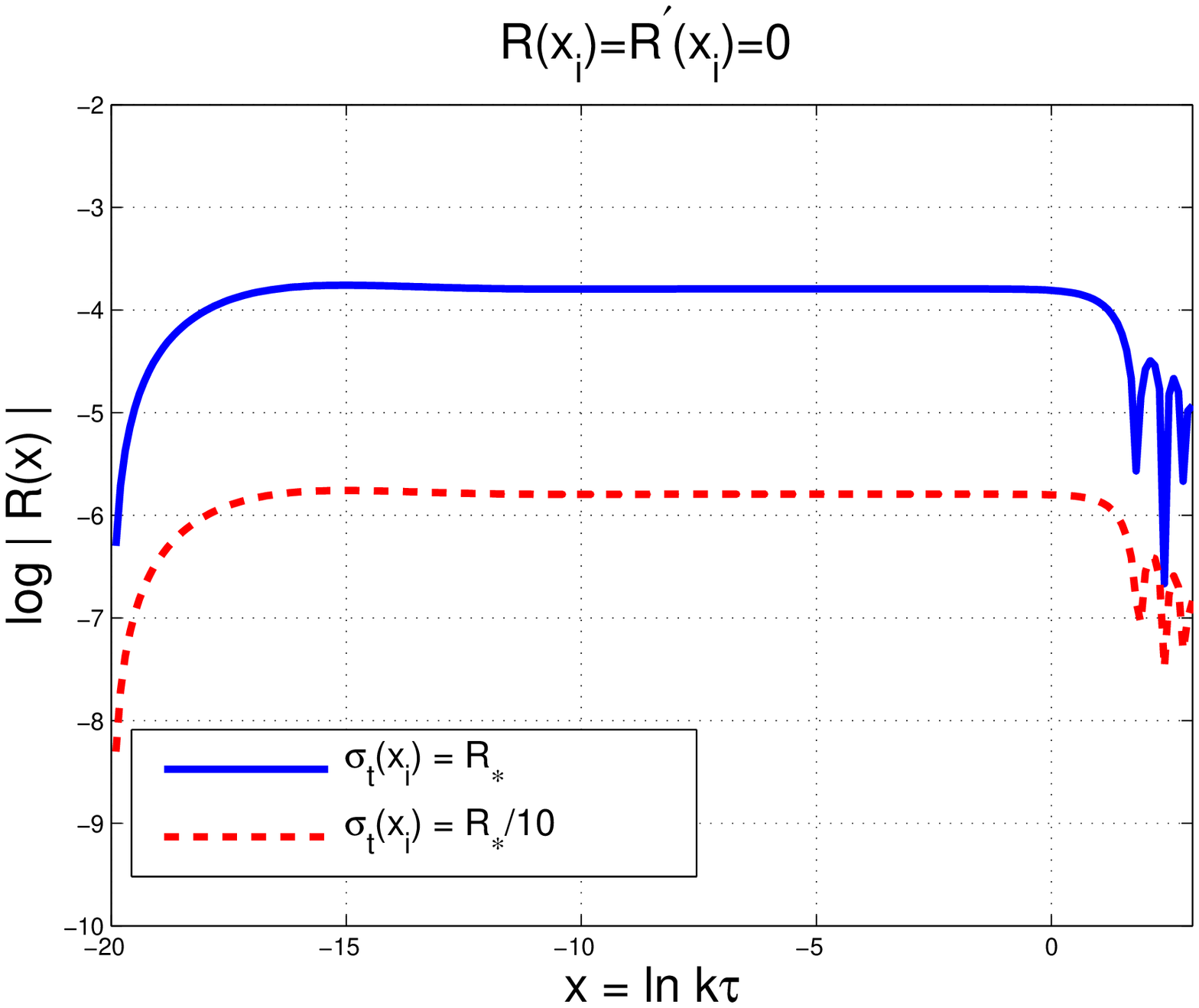}
\includegraphics[height=6.8cm]{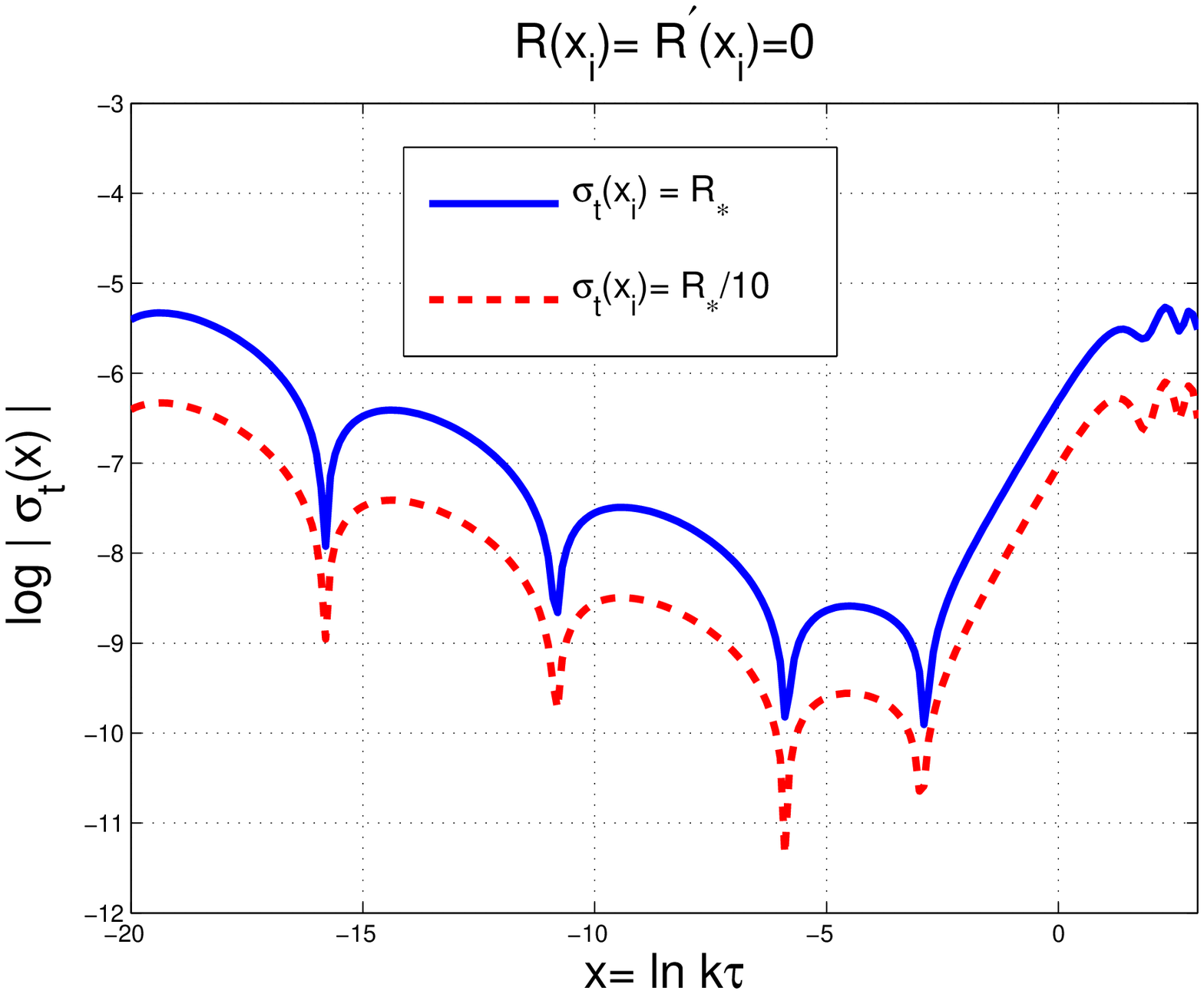}
\caption[a]{Illustrative examples where the initial conditions are dominated by the total anisotropic stress without leading 
to an early departure from the linear regime as suggested by Fig. \ref{Figure2}.}
\label{Figure3}      
\end{figure}

Specific examples confirm the conclusions of Fig. \ref{Figure2} but different examples infirm them. 
For instance the results of Fig. \ref{Figure3} are, apparently, in sharp contrast 
with the ones of Fig. \ref{Figure2}. In Fig. \ref{Figure3}, initially, ${\mathcal R}(x_{i}) = {\mathcal R}'(x_{i}) =0$ while $\sigma_{\mathrm{t}}(x_{i}) \neq 0$.
The full and dashed lines in Fig. \ref{Figure3} correspond to different initial values of $\sigma_{\mathrm{t}}(x_{i})$ which are 
given, for comparison with Figs. \ref{Figure1} and \ref{Figure2}, in units of ${\mathcal R}_{*}$.

By looking at the left and right panels of Fig. \ref{Figure3} we clearly see that both ${\mathcal R}$ and $\sigma_{\mathrm{t}}$ 
are bounded and do not diverge. This is in contrast with what happens in the example of Eq. (\ref{Figure2}) where the initial 
conditions are dominated, as in Figure \ref{Figure3}, by the anisotropic stress. 
The examples of Fig. \ref{Figure3} have been constructed by using some of the analytic results
that we are going to present in the remaining part of this section. In particular they hold provided ${\mathcal R}'(x_{i})$ and 
$\sigma_{t}(x_{i})$ depend on a fixed combination of $\sigma_{t}'(x_{i})$ and $\sigma_{t}''(x_{i})$. For the moment the apparent disagreement
 between Figs. \ref{Figure2} and \ref{Figure3} can be taken as suggestive of the limitations of a purely numerical analysis. 

\subsection{The space of the initial Cauchy data}
To solve the problem posed by Figs. \ref{Figure1},  \ref{Figure2} and \ref{Figure3}
 it is interesting to derive more general criteria able to provide sufficient conditions for the avoidance 
of the sharp growth of the curvature perturbations when the Cauchy data are dominated 
by the anisotropic stress.  The first step along this direction is to express Eqs. (\ref{ANN6a}) and (\ref{ANN6}) 
in terms of an appropriately shifted variable defined as $w = ( x - x_{i})$. The initial  conditions at $x = x_{i}$ can then be translated into Cauchy data 
for $w=0$. In terms of $w$ Eqs. (\ref{ANN6a}) and (\ref{ANN6}) can the  be expressed as:
\begin{eqnarray}
\frac{d^{3}\sigma_{\mathrm{t}}}{d w^3} - 3 \frac{d^{2} \sigma_{\mathrm{t}}}{dw^2} + \biggl( \frac{8}{5}  R_{\nu}+ 2 + \frac{6}{7} e^{2 (w + x_{i})} \biggr) \frac{d \sigma_{\mathrm{t}}}{d w} 
&=& 16 R_{\nu} \sigma_{\mathrm{t}} - \frac{48 R_{\nu}}{5} \biggl(\frac{ d{\mathcal R}}{d w} + \frac{{\mathcal R}}{9} e^{2 (w + x_{i})}\biggr) 
\label{ANN6ba}\\
\frac{d^2 {\mathcal R}}{d w^2} + \frac{d {\mathcal R}}{d w}  + c_{\mathrm{st}}^2 \,e^{ 2 (w + x_{i})} {\mathcal R} &=&  \biggl( \frac{d\sigma_{\mathrm{t}}}{d w} + 2 \sigma_{\mathrm{t}} \biggr).
\label{ANN6b}
\end{eqnarray}
Taking now the Laplace transform of both sides of Eq. (\ref{ANN6a}) we obtain the following 
difference equation:
\begin{equation}
g(s) s (s + 1) - (s+2) f(s) - (1 + s) {\mathcal R}(0) + \sigma_{\mathrm{t}}(0) - {\mathcal R}'(0) + c_{\mathrm{st}}^2 e^{2 x_{i}} g(s-2) =0,
\label{LT1}
\end{equation}
where $g(s)$ and $f(s)$ are, respectively, the Laplace transforms of ${\mathcal R}(w)$ 
and of $\sigma_{\mathrm{t}}(w)$. Taking then the Laplace transform of both sides of Eq. (\ref{ANN6b}), the following 
equation is obtained:
\begin{eqnarray}
&& [ 8 R_{\nu} ( s -10) + 5 s (s-1) (s-2) ]\, f(s) + 48 R_{\nu} \, s \, g(s) - 48 R_{\nu} {\mathcal R}(0) 
\nonumber\\
&& - [ 10 + 8 R_{\nu}  + 5 s (s -3)] \sigma_{\mathrm{t}}(0)  - 5 (s -3) \sigma_{\mathrm{t}}'(0) - 5 \sigma_{\mathrm{t}}''(0)  + 
\nonumber\\
&& 10 e^{ 2 x_{i}} \biggl\{ \frac{3}{7} [ (s-2) f(s-2) - \sigma_{\mathrm{t}}(0)] + \frac{8}{15} R_{\nu} g(s-2) \biggr\} =0.
\label{LT2}
\end{eqnarray}
\begin{figure}[!ht]
\centering
\includegraphics[height=6.8cm]{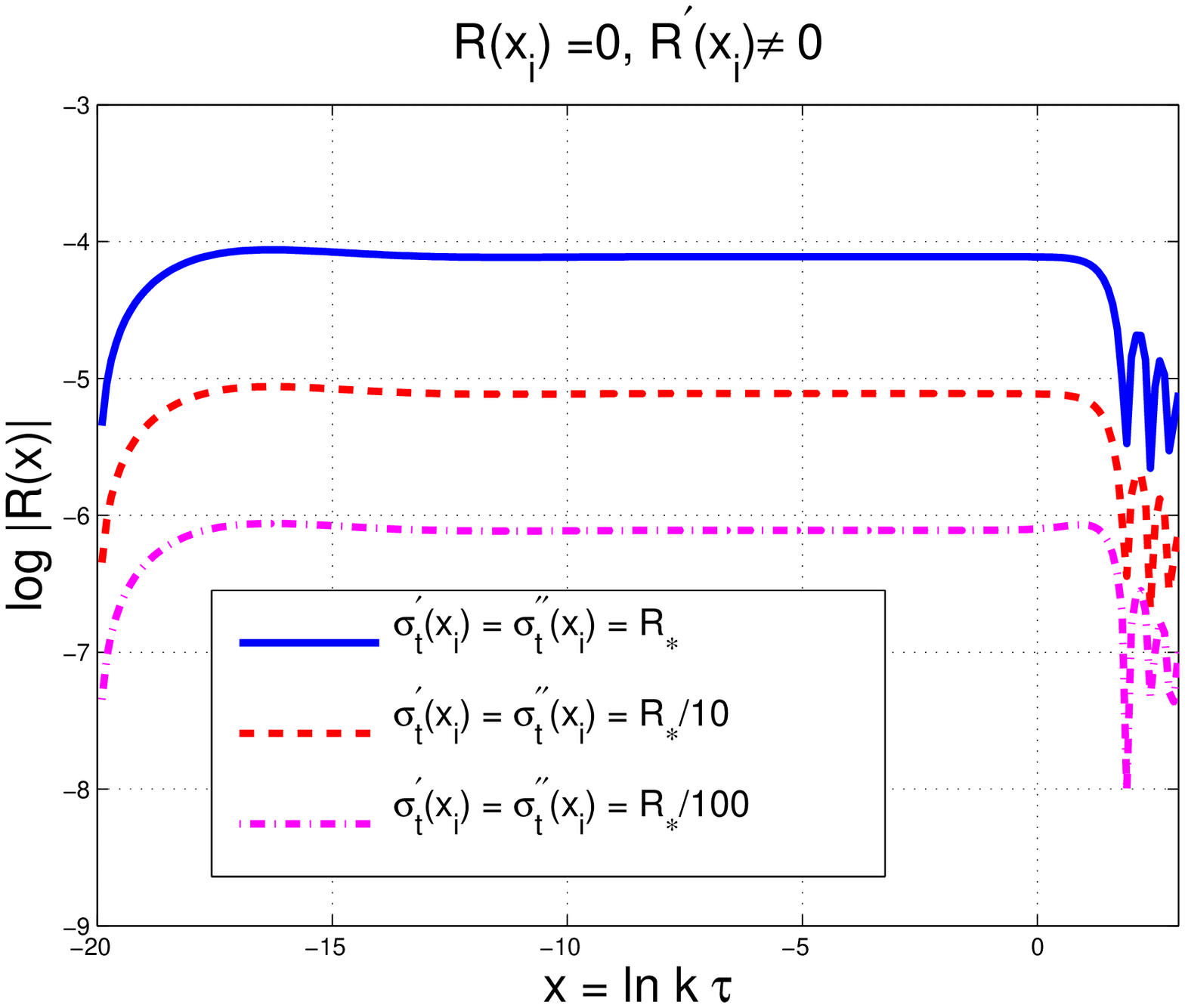}
\includegraphics[height=6.8cm]{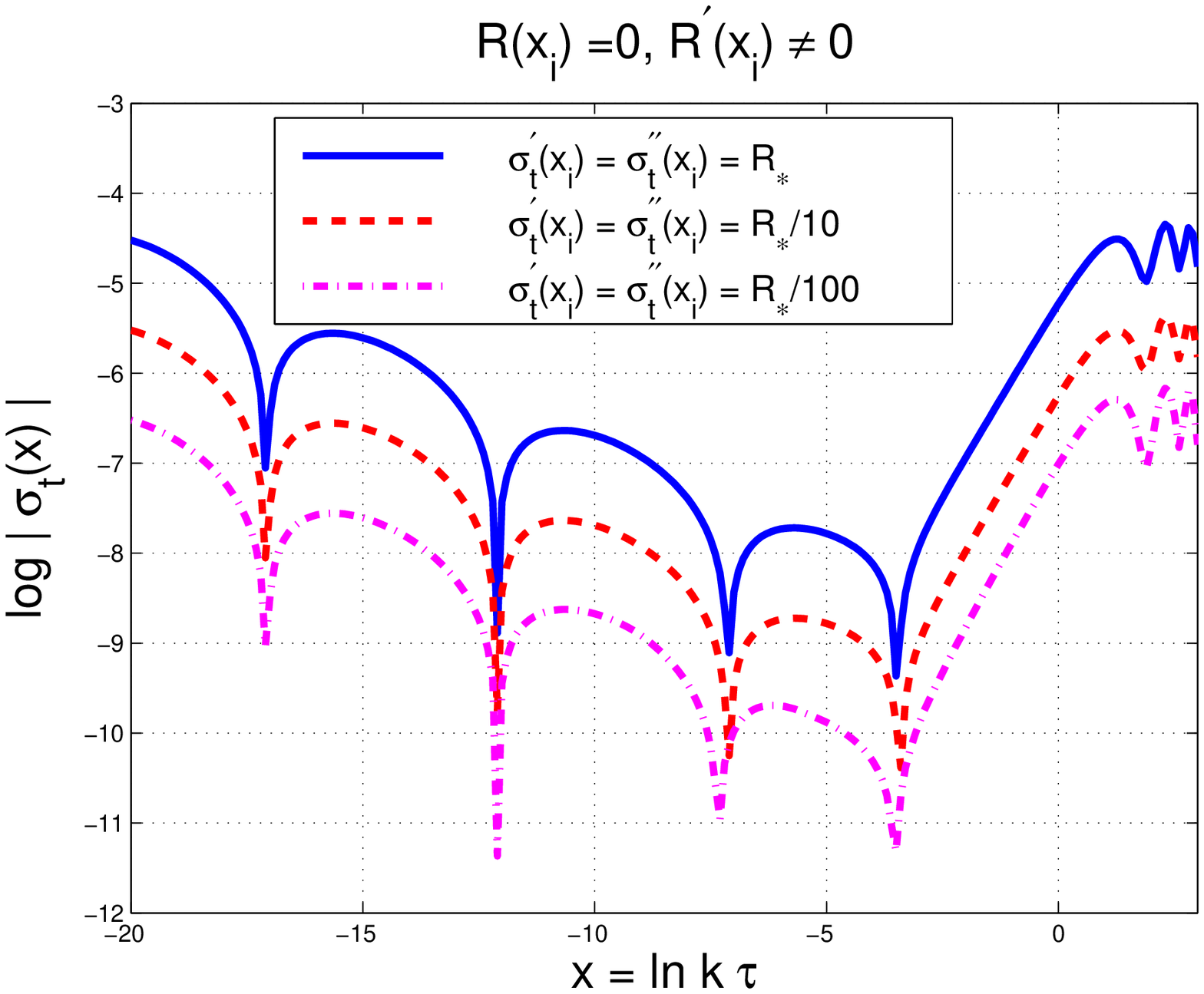}
\caption[a]{Examples of Cauchy data dominated by the anisotropic stress where 
initially ${\mathcal R}(x_{i}) \neq {\mathcal R}'(x_{i})$. The initial conditions are fixed in compliance with Eqs. (\ref{LT15a}). }
\label{Figure4}      
\end{figure}

Equations (\ref{LT1}) and (\ref{LT2}) form a system of difference equations 
that can be solved iteratively by neglecting, in the first approximation, the terms 
containing $\exp{(2 x_{i})}$. As already mentioned earlier on in this section, 
$x_{i} = \ln{(k \tau_{i})}$ is relatively large and negative for the physical set of initial conditions. Typically 
$x_{i} \simeq {\mathcal O}(-20)$. Equation (\ref{LT1}) can thus be written as:
\begin{equation}
 g(s) s (s + 1) - (s+2) f(s) = (1 + s) {\mathcal R}(0) - \sigma_{\mathrm{t}}(0) + {\mathcal R}'(0),
\label{LT1a}
\end{equation}
while, under the same approximation, Eq. (\ref{LT2}) becomes:
\begin{eqnarray}
  48 R_{\nu} \, s \,g(s) &+& [ 8 R_{\nu} ( s -10) + 5 s (s-1) (s-2) ] f(s) = 48 R_{\nu} {\mathcal R}(0) 
\nonumber\\
&+&  [ 10 + 8 R_{\nu}  + 5 s (s -3)] \sigma_{\mathrm{t}}(0) + 5 (s -3) \sigma_{\mathrm{t}}'(0) + 5 \sigma_{\mathrm{t}}''(0).
\label{LT2a}
\end{eqnarray}
The explicit solution of Eqs. (\ref{LT1a}) and (\ref{LT2a})  is 
\begin{eqnarray}
g(s) &=& a_{1}(s) {\mathcal R}(0) + a_{2}(s) {\mathcal R}'(0) + a_{3}(s) \sigma_{\mathrm{t}}(0) + a_{4}(s) \sigma_{\mathrm{t}}'(0) + 
a_{5}(s) \sigma_{\mathrm{t}}''(0),
\label{LT3}\\
f(s) &=& b_{1}(s) \sigma_{\mathrm{t}}(0) + b_{2}(s) \sigma_{\mathrm{t}}'(0) + b_{3}(s)\sigma_{\mathrm{t}}''(0) + b_{4}(s) {\mathcal R}'(0).
\label{LT4}
\end{eqnarray}
The functions $a_{j}(s)$ (with $j = 1,\,2,\,3,\,4,\,5$) multiplying the initial conditions in Eqs. (\ref{LT3}) are given by:
\begin{eqnarray}
&& a_{1}(s) = \frac{1}{s}, \qquad a_{2}(s) = \frac{4[8 R_{\nu}\, (s -10) + 5 s (s-1) (s-2)]}{5\,s \, (s -1)\, (s-2) [ (2 s +1)^2 + \beta^2]},
\nonumber\\
&& a_{3}(s) = \frac{4[ 96 R_{\nu} + 10 (s-1) (s-2)]}{5 \,s \, (s -1)\, (s-2) [ (2 s +1)^2 + \beta^2]},
\nonumber\\
&& a_{4}(s) = \frac{4 (s + 2) (s - 3)}{s ( s- 1) (s -2) [ (2 s +1)^2 + \beta^2]},
\nonumber\\
&& a_{5}(s) = \frac{4 (s + 2) }{s ( s- 1) (s -2) [ (2 s +1)^2 + \beta^2]},
\label{LT6}
\end{eqnarray}
where we introduced the auxiliary quantity\footnote{Notice that $\beta(k,\tau)$ denoted the off-diagonal component of the metric appearing in section \ref{sec4} but no confusion is possible since $\beta(k,\tau)$ and $\beta(R_{\nu})$ never appear in the same context. } $\beta(R_{\nu}) = \sqrt{32 R_{\nu}/5 -1}$; 
With the same notations the functions $b_{i}(s)$ (with $i = 1,\,2,\,3,\,4$) appearing in Eq. (\ref{LT4}) are instead given by:
\begin{eqnarray}
&& b_{1}(s) = \frac{4[ 5 (s + 1) (s -1) (s-2) + 8 R_{\nu} (s + 7)]}{ 5 (s -1 ) (s - 2) [ (2 s +1)^2 + \beta^2]},
\nonumber\\
&& b_{2}(s) = \frac{4 (s+1) (s - 3)}{ (s -1 ) (s - 2) [ (2 s +1)^2 + \beta^2]},
\nonumber\\
&& b_{3}(s) = \frac{4 (s +1)}{(s-1) (s - 2)  [ (2 s +1)^2 + \beta^2]},
\nonumber\\
&& b_{4}(s) = - \frac{192 \, R_{\nu}}{5 (s-1) (s -2)  [ (2 s +1)^2 + \beta^2]}.
\label{LT8}
\end{eqnarray}

We can finally take the inverse transform of Eqs. (\ref{LT3}) and (\ref{LT4}) and go back either to the $w$ variable or to the $x$ variable. 
Except $a_{1}(s)$ (whose antitransform is constant in $w$-space), the structure of the poles of Eqs. (\ref{LT6}) and (\ref{LT8})
implies that each of the $a_{j}(s)$ and $b_{i}(s)$ leads, generally speaking, to three qualitatively different contributions:
\begin{itemize}
\item{} two exponential contributions increasing, respectively, as $e^{2 w}$ and as $e^{w}$: these 
are the most dangerous terms that may take the curvature perturbations far from the linear regime 
at early time;
\item{} a constant contribution possibly depending on $R_{\nu}$;
\item{} an exponentially suppressed contribution going as $e^{-w/2}$; this contribution is modulated by linear combinations 
oscillating factors going proportional to $\cos{(\beta w/2)}$ and to $\sin{(\beta w/2)}$.
\end{itemize}
For instance the anti-transform of $a_{2}(s)$ can be written, in $w$-space as: 
\begin{eqnarray}
 {\mathcal L}^{-1}[ a_{2}(s)] &=& - \frac{128 R_{\nu}}{5 ( 25 + \beta^2)} e^{2 w} 
+ \frac{288 R_{\nu}}{5 (9 + \beta^2)} e^{w}  - \frac{32 R_{\nu}}{1 + \beta^2} + e^{- w/2} j(w,\beta),
\nonumber\\
j(w,\beta) &=& \frac{768 \left(\beta^2-39\right) R_{\nu} }{5 \beta  \left(\beta ^2+1\right) \left(\beta ^2+9\right) \left(\beta ^2+25\right)}\,\cos{(\beta w/2})
\nonumber\\
&+& 2 \biggl[ \frac{1}{\beta} - \frac{32 \left(\beta ^4+166 \beta ^2-315\right) R_{\nu} }{5 \beta  \left(\beta ^2+1\right) \left(\beta ^2+9\right) \left(\beta ^2+25\right)}\biggr] \sin{(\beta w/2)}.
\nonumber
\end{eqnarray}
\begin{figure}[!ht]
\centering
\includegraphics[height=6.8cm]{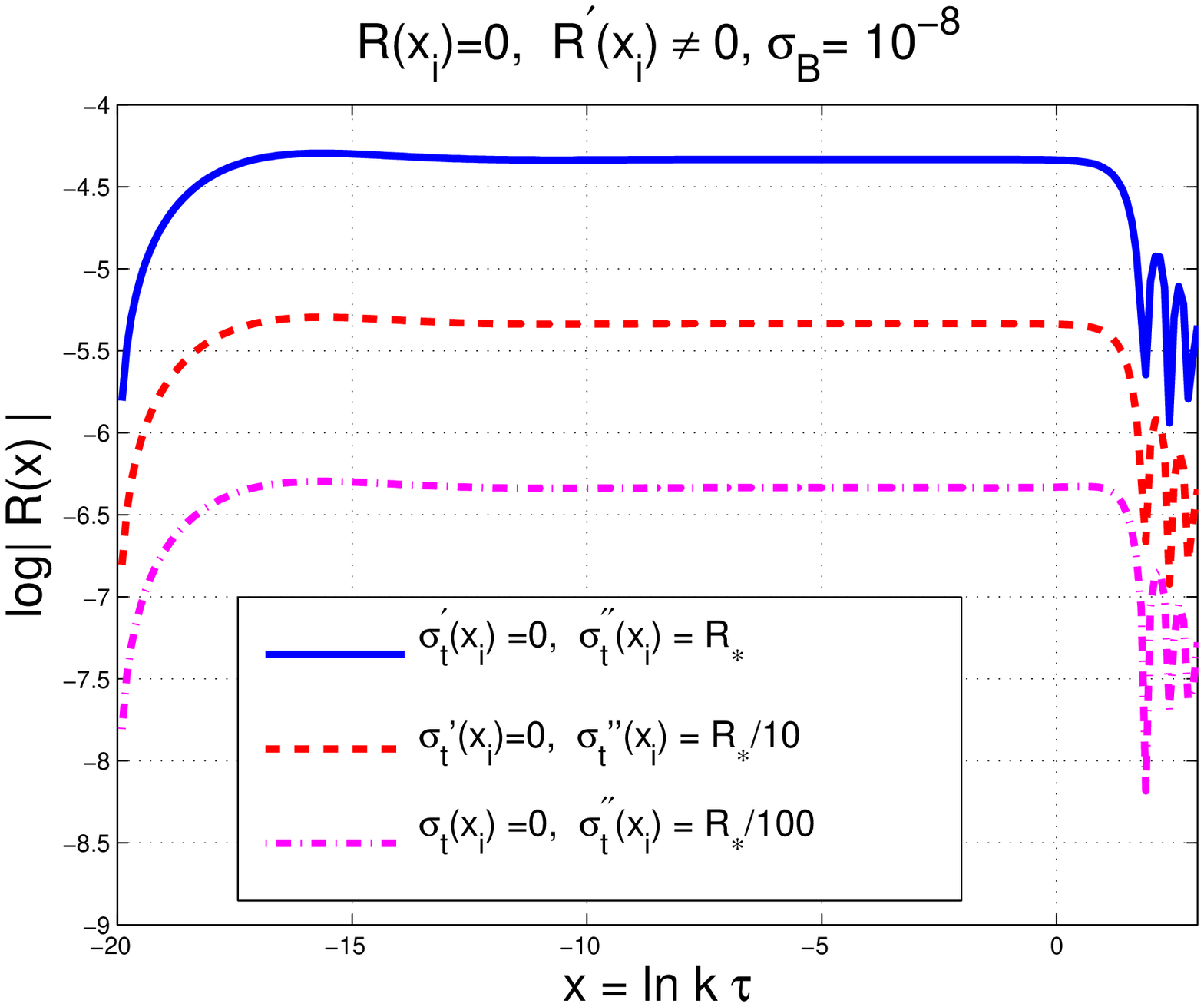}
\includegraphics[height=6.8cm]{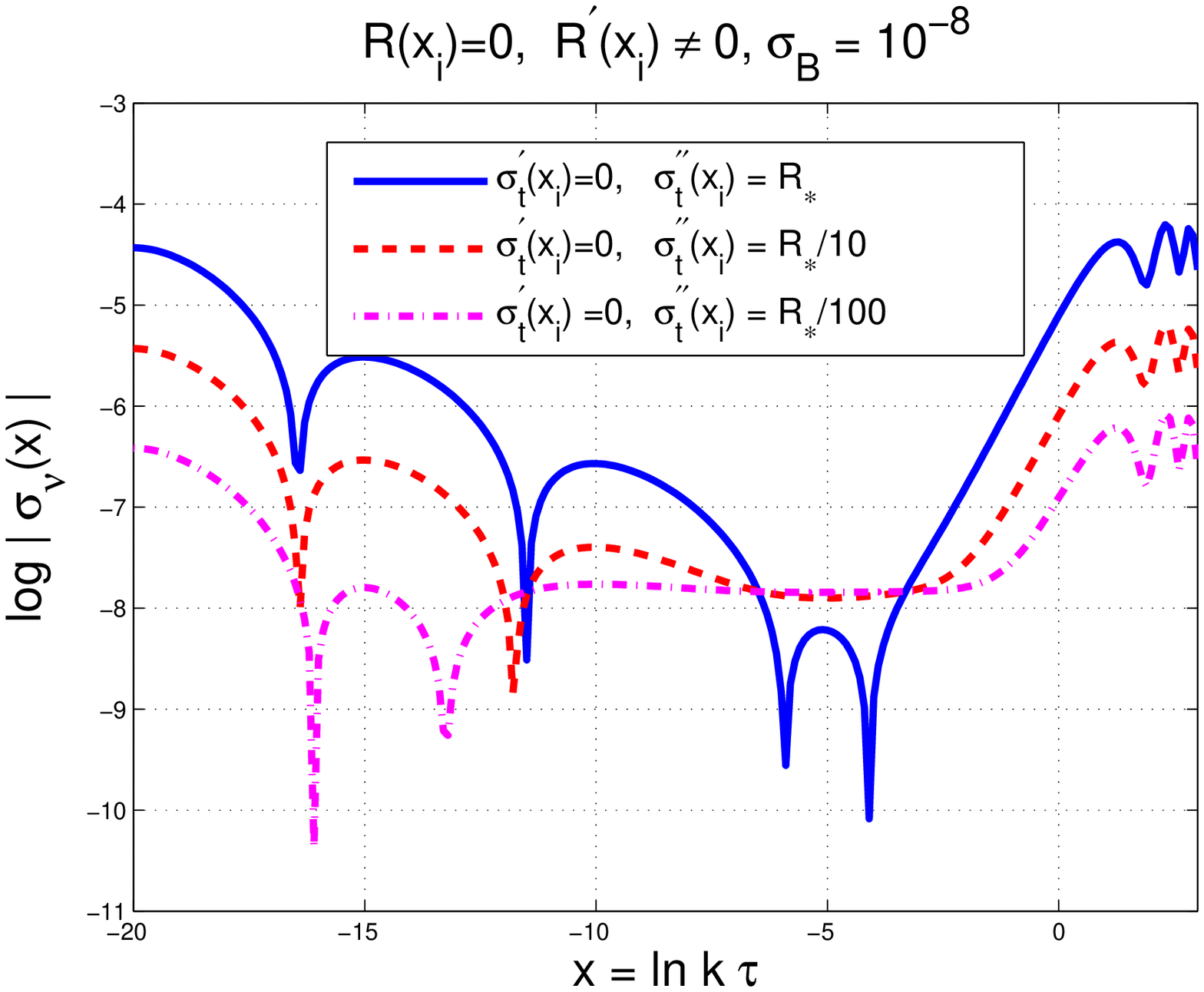}
\caption[a]{Illustrative examples of asymmetric initial conditions where $\sigma_{\mathrm{t}}(x_{i}) \neq 0$ and 
$\sigma_{\mathrm{t}}'(x_{i}) \neq \sigma_{\mathrm{t}}''(x_{i})$. As in Fig. \ref{Figure4} the Cauchy data are fixed according to Eqs. (\ref{LT15a}).}
\label{Figure5}      
\end{figure}

Being exponentially suppressed, the oscillating contribution is immaterial for our discussion. Consequently the inverse Laplace transform of Eqs. (\ref{LT3}) and (\ref{LT4}) gives, in the $x$-parametrization:
\begin{eqnarray}
{\mathcal R}(x) &=& {\mathcal R}_{*} + \biggl[ - \frac{16 R_{\nu}}{15 + 4 R_{\nu}} e^{ 2 (x - x_{i})}+ 
\frac{36 R_{\nu}}{5 + 4 R_{\nu}} e^{(x - x_{i})}\biggr] {\mathcal R}'(x_{i})
\nonumber\\
&+& \biggl[ \frac{24 R_{\nu}}{15 + 4 R_{\nu}} e^{ 2 ( x - x_{i})} - \frac{48 R_{\nu}}{5 + 4 R_{\nu}} e^{(x - x_{i})} \biggr] \sigma_{\mathrm{t}}(x_{i}) 
\nonumber\\
&+&
 \biggl[ -\frac{5}{15 + 4 R_{\nu}} e^{ 2 ( x - x_{i})} + \frac{15}{5 + 4 R_{\nu}} e^{(x - x_{i})} \biggr] \sigma_{\mathrm{t}}'(x_{i}) 
 \nonumber\\
 &+&  \biggl[ \frac{5}{15 + 4 R_{\nu}} e^{ 2 ( x - x_{i})} - \frac{15}{2(5+ 4 R_{\nu})} e^{(x - x_{i})} \biggr] \sigma_{\mathrm{t}}''(x_{i}) + {\mathcal O}\biggl(e^{- (x -x_{i})/2}\biggr),
\label{LT9}\\
\sigma_{\mathrm{t}}(x) &=& \biggl[ \frac{36 R_{\nu}}{15 + 4 R_{\nu}} e^{2( x- x_{i})} - \frac{32 R_{\nu}}{5 + 4 R_{\nu}} e^{(x- x_{i})}\biggr] 
\sigma_{\mathrm{t}}(x_{i}) 
\nonumber\\
&+& \biggl[ -\frac{15}{2(15 + 4 R_{\nu})} e^{2( x- x_{i})} + \frac{10}{5 + 4 R_{\nu}} e^{(x- x_{i})}\biggr] \sigma_{\mathrm{t}}'(x_{i}) 
\nonumber\\
 &+& \biggl[ \frac{15}{2(15 + 4 R_{\nu})} e^{2( x- x_{i})} - \frac{5}{5 + 4 R_{\nu}} e^{(x- x_{i})}\biggr] \sigma_{\mathrm{t}}''(x_{i})
 \nonumber\\
 &+&   \biggl[- \frac{24 R_{\nu}}{(15 + 4 R_{\nu})} e^{2( x- x_{i})} +\frac{24 R_{\nu}}{5 + 4 R_{\nu}} e^{(x- x_{i})}\biggr] {\mathcal R}'(x_{i}) + {\mathcal O}\biggl(e^{- (x -x_{i})/2}\biggr).
 \label{LT10}
 \end{eqnarray}

\subsection{Taming the early departures from the linear regime}
The presence (or the absence) of the diverging contributions in Eqs. (\ref{LT9}) and (\ref{LT10}) depends upon the 
Cauchy data.  The exponentially increasing terms disappear from Eqs. (\ref{LT9}) and (\ref{LT10}) provided a pair of specific relations 
connects ${\mathcal R}'(x_{i})$ and $\sigma_{t}(x_{i})$ to $ \sigma_{\mathrm{t}}'(x_{i})$ and $\sigma_{\mathrm{t}}''(x_{i})$.

More precisely, by looking at the leading contributions of  Eq. (\ref{LT9}) we can demand that the coefficients of the exponentially increasing contributions vanish identically for the initial Cauchy data. This request leads to a pair of relations among the initial data:
\begin{eqnarray}
&& 8 R_{\nu} [ 3 \sigma_{\mathrm{t}}(x_{i}) - 2 {\mathcal R}'(x_{i}) ] - 
5 [\sigma_{\mathrm{t}}'(x_{i}) -\sigma_{\mathrm{t}}''(x_{i})] =0,
\label{LT11}\\
&& 24 R_{\nu}[3  {\mathcal R}'(x_{i}) - 4 \sigma_{\mathrm{t}}(x_{i})]  +15 [ 2 \sigma_{\mathrm{t}}'(x_{i}) -  \sigma_{\mathrm{t}}''(x_{i}) ]=0.
\label{LT12}
\end{eqnarray}
The results of Eqs. (\ref{LT11}) and (\ref{LT12}) justify and explain the examples of Fig. \ref{Figure3}: the initial conditions 
assigned in the numerical integration leading to Fig. \ref{Figure3} have been selected in such a way 
that Eqs. (\ref{LT11}) and (\ref{LT12}) are satisfied.  

The same strategy can be used in Eq. (\ref{LT10}) and we can demand that the coefficients of the exponentially increasing contributions vanish identically for the initial Cauchy 
data. The two supplementary relations obtained with this procedure are:
\begin{eqnarray}
&& 24 R_{\nu}[ 3 \sigma_{\mathrm{t}}(x_{i})  - 2 {\mathcal R}'(x_{i})] -  15[ \sigma_{\mathrm{t}}'(x_{i}) -  \sigma_{\mathrm{t}}''(x_{i})] =0,
\label{LT13}\\
&& 8 R_{\nu}[ 3 {\mathcal R}'(x_{i})- 4\sigma_{\mathrm{t}}(x_{i})] + 5 [ 2 \sigma_{\mathrm{t}}'(x_{i}) - \sigma_{\mathrm{t}}''(x_{i})] =0. 
\label{LT14}
\end{eqnarray}
The system formed by Eqs. (\ref{LT11})--(\ref{LT12}) is not  linearly independent from the system of Eqs. (\ref{LT13})--(\ref{LT14}): if we multiply by a factor 
of $3$ Eq. (\ref{LT11}) we obtain Eq. (\ref{LT13}) and if we multiply by a factor of $3$ Eq. (\ref{LT14}) we obtain Eq. (\ref{LT12}). Thus,
if Eqs. (\ref{LT11}) and (\ref{LT12}) are satisfied also Eqs. (\ref{LT13})--(\ref{LT14}) will also be satisfied. 

\begin{figure}[!ht]
\centering
\includegraphics[height=6.8cm]{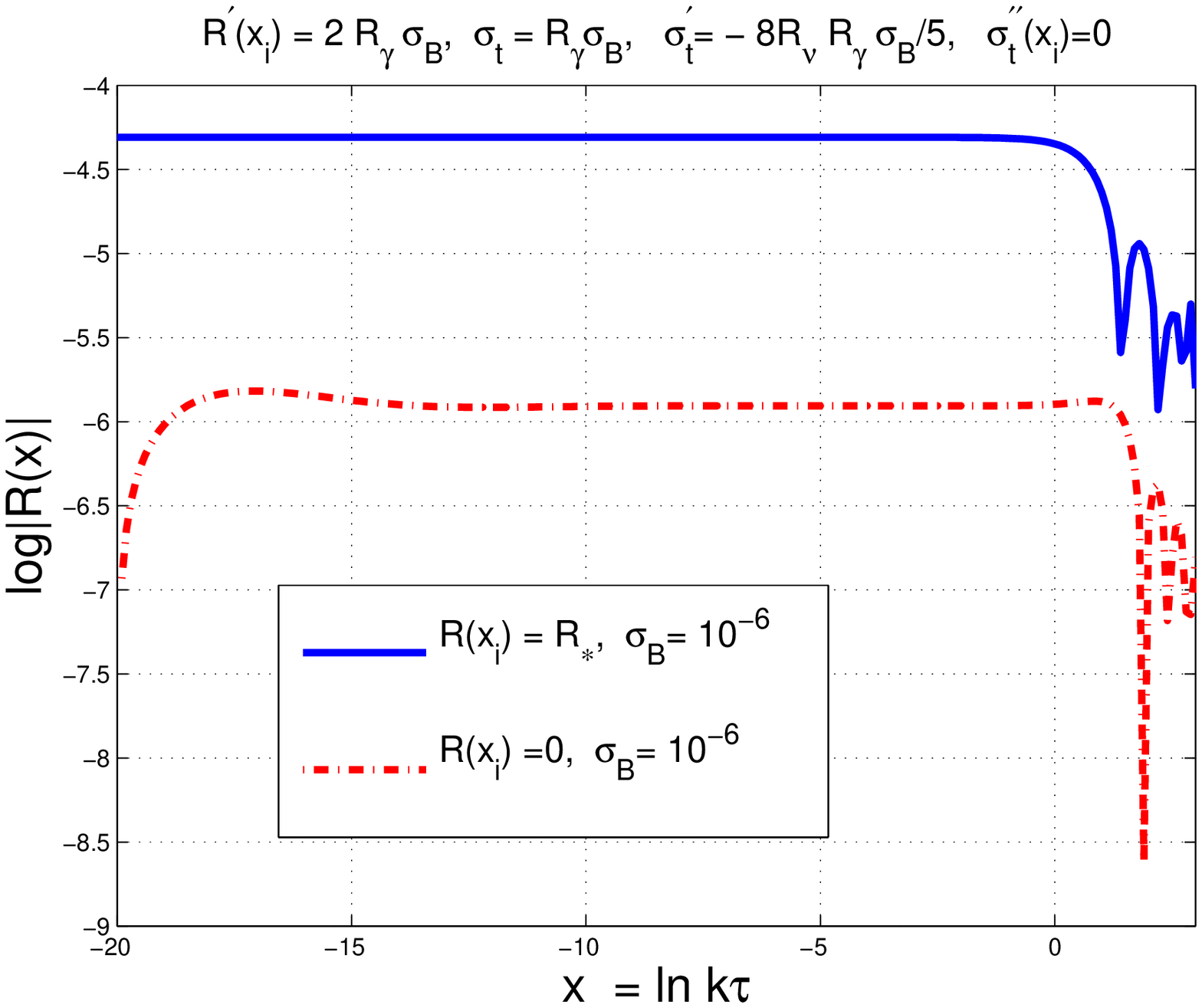}
\includegraphics[height=6.8cm]{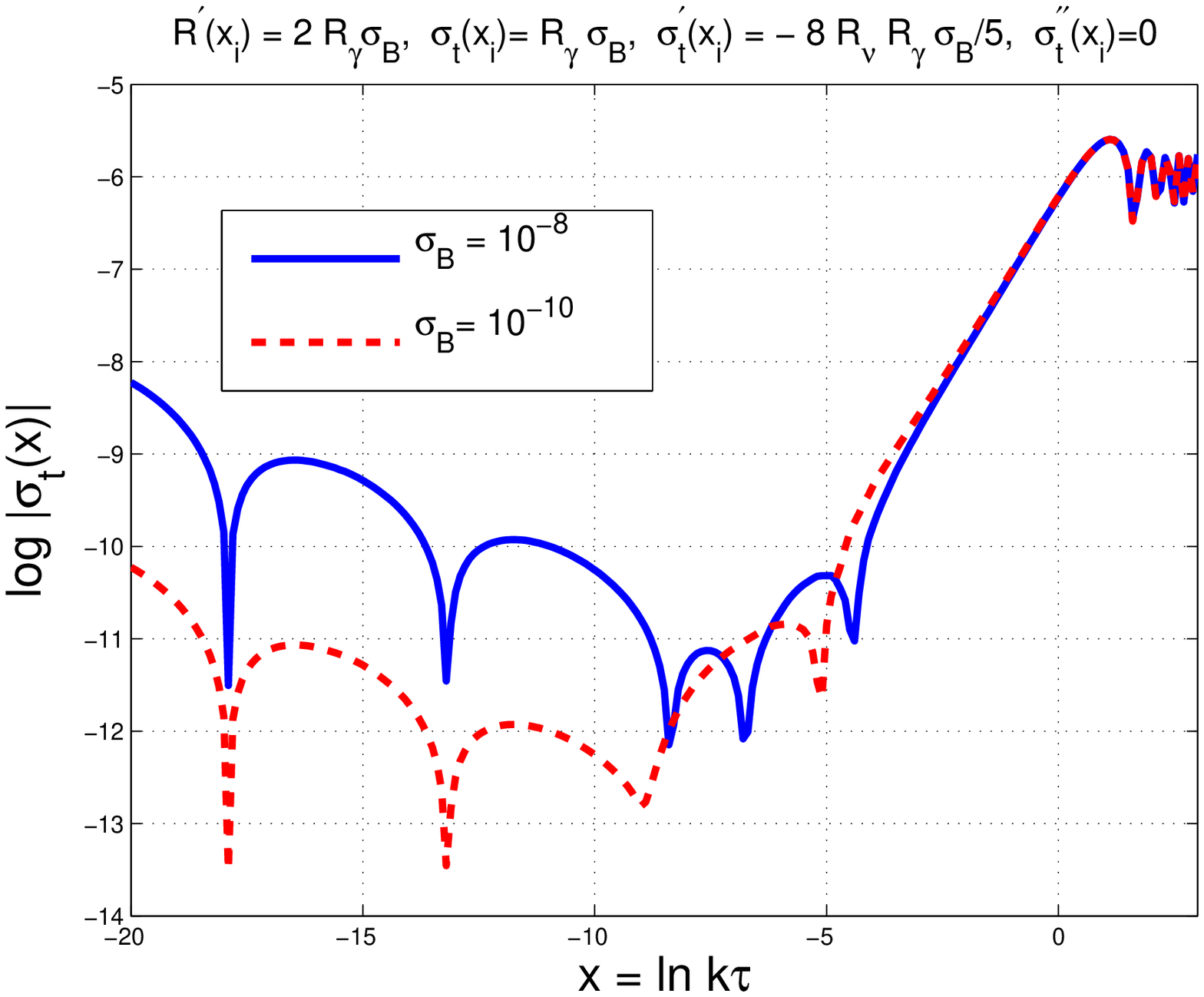}
\caption[a]{The numerical integration corresponding to the Cauchy data of Eqs. (\ref{M1}) and (\ref{M2}). As in Figs. \ref{Figure4}  and \ref{Figure5} the Cauchy data are fixed according to Eqs. (\ref{LT15a}).}
\label{Figure6}      
\end{figure}

Equations (\ref{LT11})--(\ref{LT12}) can be solved by relating the derivatives of the anisotropic stress 
to $\sigma_{\mathrm{t}}(x_{i})$ and to ${\mathcal R}'(x_{i})$:
\begin{equation}
\sigma_{\mathrm{t}}'(x_{i}) = - \frac{8}{5} R_{\nu} [ {\mathcal R}'(x_{i}) - \sigma_{\mathrm{t}}(x_{i})] , \qquad \sigma_{t}''(x_{i}) = \frac{8}{5}R_{\nu}  [ {\mathcal R}'(x_{i}) - 2\sigma_{\mathrm{t}}(x_{i})].
\label{LT15}
\end{equation}
Equation (\ref{LT15}) provides a quantitative solution to the problem posed by Figs. \ref{Figure2} and \ref{Figure3}.  The first and second derivatives of the anisotropic stress must be tuned in order to tame the exponential 
growth of the curvature perturbations for a generic initial value of ${\mathcal R}'$ and $\sigma_{t}$. The inverse 
viewpoint can be adopted by using  $\sigma_{\mathrm{t}}'(x_{i})$ and $\sigma_{\mathrm{t}}''(x_{i})$ as pivotal variables. 
Indeed, inverting Eq. (\ref{LT15}) we will have:
\begin{equation}
{\mathcal R}'(x_{i}) = - \frac{5}{8 R_{\nu}} [2 \sigma_{\mathrm{t}}'(x_{i}) + \sigma_{\mathrm{t}}''(x_{i})],\qquad \sigma_{\mathrm{t}}(x_{i}) = 
- \frac{5}{8 R_{\nu}} [\sigma_{\mathrm{t}}'(x_{i}) + \sigma_{\mathrm{t}}''(x_{i})].
\label{LT15a}
\end{equation}
The criteria given by Eq. (\ref{LT15}) (or equivalently by Eq. (\ref{LT15a})) have been obtained after 
various steps involving some plausible approximations. It is therefore interesting to present direct
numerical tests of these criteria. The illustrative examples of Figs. \ref{Figure4}, \ref{Figure5} and \ref{Figure6}
confirm the heuristic validity of the relations (\ref{LT15}) and (\ref{LT15a}).

In Fig. \ref{Figure4} we explore a set of initial data dominated by the anisotropic 
stress but different from the initial conditions leading to Fig. \ref{Figure3}. While 
 ${\mathcal R}(x_{i}) = {\mathcal R}'(x_{i}) =0$ in Fig. \ref{Figure3}, in Fig. \ref{Figure4}
 ${\mathcal R}(x_{i}) =0$ and  ${\mathcal R}'(x_{i}) \neq 0$. The values of ${\mathcal R}'(x_{i})$ 
 and of $\sigma_{\mathrm{t}}(x_{i})$ are fixed, in terms of $\sigma_{\mathrm{t}}'(x_{i})$ 
 and of $\sigma_{\mathrm{t}}''(x_{i})$ by Eq. (\ref{LT15a}). In the legend of Fig. \ref{Figure4} 
 the initial values of $\sigma_{\mathrm{t}}'(x)$ 
 and of $\sigma_{\mathrm{t}}''(x)$ are explicitly mentioned.
 As implied by the criteria leading to Eq. (\ref{LT15a}) we do expect the absence of diverging 
 contributions of the type of Fig. \ref{Figure2}. This is exactly what happens. Notice that
 none of the fluctuations illustrated in Fig. \ref{Figure4} become nonlinear while 
 the relevant wavelengths are still larger than the Hubble radius.

In Fig. \ref{Figure5} we explore yet a different set of initial conditions that are asymmetric both in the 
curvature and in the anisotropic stress. In the left panel we illustrate, as usual, the evolution 
of the curvature perturbations for different values of $\sigma_{\mathrm{t}}''(x_{i})$ 
(expressed in units of ${\mathcal R}_{*}$). In the right panel of Fig. \ref{Figure5}
we illustrate directly the anisotropic stress 
of the neutrinos (instead of the total anisotropic stress containing also the magnetic contribution). 
 As in the case of Fig. \ref{Figure4}  the values of ${\mathcal R}'(x_{i})$ 
 and of $\sigma_{\mathrm{t}}(x_{i})$ are fixed, in terms of $\sigma_{\mathrm{t}}'(x_{i})$ 
 and of $\sigma_{\mathrm{t}}''(x_{i})$ by Eq. (\ref{LT15a}). Also in Fig. \ref{Figure5} the criteria expressed by Eqs. (\ref{LT15}) and (\ref{LT15a}) are clearly verified. 

As a last example let us consider the results of Fig. \ref{Figure6}. If  for $x= x_{i}$ the only contribution to the total anisotropic stress 
is given only by the magnetic term, then $\sigma_{\mathrm{t}} = R_{\gamma} \sigma_{B}$ but $\sigma_{\nu}(x_{i}) =0$.
Recalling the solution obtained in Eq. (\ref{ANN21})
we shall have that ${\mathcal R}(x) = {\mathcal R}_{*} + 2 R_{\gamma} \sigma_{B} (x - x_{i})$, implying that 
\begin{equation}
{\mathcal R}(x_{i}) = {\mathcal R}_{*}, \qquad {\mathcal R}'(x_{i}) = 2 R_{\gamma} \sigma_{B}.
\label{M1}
\end{equation}
The fate of these initial data can be investigated using the results developed so far.
The previous analysis tells that if $\sigma_{\mathrm{t}}'(x_{i})$ and $\sigma_{\mathrm{t}}''(x_{i})$ are assigned 
arbitrarily, the curvature perturbations will become eventually nonlinear unless 
\begin{equation}
\sigma_{\mathrm{t}}^{\prime}(x_{i}) = - \frac{8}{5} R_{\nu} R_{\gamma} \sigma_{B}, \qquad 
\sigma_{\mathrm{t}}^{\prime\prime}(x_{i}) =0. 
\label{M2}
\end{equation}
The initial conditions of Eqs. (\ref{M1}) and (\ref{M2}) are located within the  tuning volume 
of the initial Cauchy data satisfying Eqs. (\ref{LT11}) and (\ref{LT12}). Equation (\ref{M2}) 
fixes the first and second derivatives of the total anisotropic stress which are not 
specified even if we already required $\sigma_{\nu}(x_{i}) =0$. Only if the Cauchy data 
satisfy Eq. (\ref{M2}) we can safely argue that the exponentially increasing contributions 
are tamed.

The initial conditions of Eqs. (\ref{M1})--(\ref{M2}) can be tested by direct numerical integration. The results are 
illustrated in Fig. \ref{Figure6} and, once more, the criteria 
derived from Eqs.  (\ref{LT11}) and (\ref{LT12}) are confirmed. In the left panel of Fig. \ref{Figure6} we compare 
the case ${\mathcal R}(x_{i}) = {\mathcal R}_{*}$ (full line) and the case ${\mathcal R}(x_{i}) =0$ (dot-dashed line). In the right panel we illustrate
the total anisotropic stress. As expected on the basis of the more general considerations 
discussed before, both the curvature perturbations and the total anisotropic stress do not exhibit 
exponentially divergent contributions that will be however present if $\sigma_{\mathrm{t}}'(x_{i})$ and $\sigma_{\mathrm{t}}^{\prime\prime}(x_{i})$ would be 
arbitrarily assigned as previously illustrated in Fig. \ref{Figure2}.
We can thus conclude that the heuristic criteria expressed by Eqs. (\ref{LT15}) and (\ref{LT15a}) are confirmed by the diverse numerical examples. 

In summary, the problem of initial data of the large-scale curvature perturbations has been reduced
to an ordinary Cauchy problem fully determined by five initial data: two involving 
the curvature perturbations and three involving the total anisotropic stress. The absence 
of exponentially increasing terms reduces the five-dimensional space to a three-dimensional 
Cauchy volume where the initial data are related by two linearly independent relations.
Our heuristic criteria have been derived by trying 
to formulate more faithfully a Cauchy problem for the large-scale curvature 
perturbations. The derived relations seem to solve a more general boundary value 
problem insofar as they give a sufficient condition for the constancy of the 
curvature perturbations even when the large-scale initial conditions are dominated 
by the anisotropic stress\footnote{We remark that the techniques discussed here can be easily 
applied to the case $\delta p_{\mathrm{nad}} \neq 0$ since all the equations 
have been derived in general terms. We refrain from doing this for reasons 
of opportunity and to avoid lengthy discussions.}.

\renewcommand{\theequation}{7.\arabic{equation}}
\setcounter{equation}{0}
\section{Concluding remarks}
\label{sec7}
We demonstrated that two coupled differential equations of second and third order describe, respectively, the evolution 
of the quasinormal mode of the plasma and of the anisotropic stress.  The total anisotropic stress, 
the non-adiabatic pressure fluctuations and the large-scale magnetic fields are present on equal footing 
as potential sources of inhomogeneities. On the basis of the reported results, the proposed approach seems reasonably simple and direct.

When all the supplementary sources of inhomogeneity vanish, the quasinormal 
modes of the plasma describe the phonons of a gravitating fluid in a conformally flat background geometry firstly 
derived by Lukash in the early eighties. The obtained system is gauge-invariant. The derivation can be presented and interpreted
both in the conformally Newtonian and in the synchronous coordinate systems. Various physical solutions 
have been obtained with particular attention to the standard concordance paradigm and to its magnetized 
completion. After solving for the quasi normal mode and for the anisotropic stress, all the scalar fluctuations of the plasma 
can be easily deduced without the addition of further assumptions. 

The large-scale symmetries of the system suggest that Cauchy data of large-scale curvature perturbations can be reduced to an initial 
value problem involving the quasinormal mode, the total anisotropic stress and its first two derivatives 
with respect to an appropriately rescaled variable measuring the ratio between the particle horizon 
and the physical wavelength. As an illustrative example, a set of consistency conditions among the Cauchy data, guaranteeing the validity of the perturbative expansion for generic anisotropic stresses, has been analytically deduced and numerically tested.

\section*{Acknowledgments}
It is a pleasure to thank T. Basaglia and A. Gentil-Beccot of the CERN scientific information service for their valuable assistance.

\end{document}